\documentclass[journal, a4paper]{IEEEtran}

\usepackage{cuted}
\usepackage{stackengine}
\usepackage{amsmath}
\def\yenrule{\rule{1.3ex}{.1ex}}
\usepackage[Algorithm]{algorithm}
\usepackage{textcomp}
\usepackage[australian]{babel}
\usepackage{ wasysym}
\usepackage{setspace}
\usepackage{graphicx}
\usepackage{multicol}
\usepackage{multirow}
\usepackage{rotating}
\usepackage[table]{xcolor} 
\usepackage{booktabs}
\usepackage{subfloat}
\usepackage{subcaption}
\usepackage{mathtools}
\usepackage{comment}
\usepackage{mathtools}
\usepackage{url}
\usepackage{tablefootnote}
\usepackage{cite}
\usepackage{amsmath,amssymb,amsfonts}
\usepackage{algorithmic}
\usepackage{graphicx}
\usepackage{textcomp}
\usepackage{xcolor}
\usepackage{lscape}
\usepackage{authblk}

\def\BibTeX{{\rm B\kern-.05em{\sc i\kern-.025em b}\kern-.08em
		T\kern-.1667em\lower.7ex\hbox{E}\kern-.125emX}}
\def\textyen{\renewcommand\stacktype{L}\stackon[.4ex]{\stackon[.65ex]{Y}{\yenrule}}{\yenrule}}

\providecommand{\keywords}[1]{\textbf{\textit{Index terms---}} #1}

\begin{document}

\title{Adaptive Scheduling for Efficient Execution of Dynamic Stream Workflows}

\author[1]{Mutaz Barika}
\author[1]{Saurabh Garg}
\author[2]{Rajiv Ranjan}
\affil[1]{University of Tasmania, Australia}
\affil[2]{Newcastle University, United Kingdom}

\maketitle

\begin{abstract}
	Stream workflow application such as online anomaly detection or online traffic monitoring, integrates multiple streaming big data applications into data analysis pipeline. This application can be highly dynamic in nature, where the data velocity may change at runtime and therefore the resources should be managed overtime. \textcolor{black}{To manage these changes, the orchestration of this application requires a dynamic execution environment and dynamic scheduling technique. For the former requirement, Multicloud environment is a visible solution to cope with the dynamic aspects of this workflow application. While for the latter requirement, dynamic scheduling technique not only need to adhere to end user's requirements in terms of data processing and deadline for decision making, and data stream sources location constraints, but also adjust provisioning and scheduling plan at runtime to cope with dynamic variations of stream data rates. Therefore,} we propose a two-phase adaptive scheduling technique to efficiently schedule dynamic workflow application in Multicloud environment that can respond to changes in the velocity of data at runtime. The experimental results showed that the proposed technique is close to the lower bound and effective for different experiment scenarios.
\end{abstract}


\section{Introduction}\label{sec:introduction}

Several IoT applications and services such as smart cities, smart parking and smart traffic control, have evolved to cope with the demand of improving our lives \cite{zanella2014internet} \cite{mehmood2017internet}. These applications are not monolithic application, but they contain a network of different analytical components which are composed in the form of a workflow to make better decisions. An example of this workflow is smart road traffic monitoring as a service of smart city services that utilizes the true power of connected vehicles in addition to roadside infrastructure (e.g. traffic lights, cameras) to create real-time view of road traffic conditions \cite{chen2017connected}. This type of workflow is also called stream workflow application and is becoming gradually viable for solving real-time data computation problems that are more complex.

\par Stream workflows are very different from traditional business and scientific workflows \cite{redlich2014research} \cite{liu2015survey} as they have to continuously process an infinite stream of data with each analytical component always in an active state. \textcolor{black}{They can be more complex and involve heterogeneity, multiple data sources and multiple outputs. They also differ from streaming operator graphs (generated by Apache Storm or Flink for example) as the source of data for the whole operator graph is one and there is one end operator, thus operator graph is just a simplified case of stream workflow.} Moreover, this workflow can be highly dynamic in nature, where the data velocity may change at runtime which reflects the load at given time, and therefore the resources should be managed overtime. \textcolor{black}{Furthermore, the use of single cloud to execute this workflow could not meet user requirements due to the distribution of external data sources, thus Multicloud environment that consolidates multiple clouds can help in utilizing data locality by orchestrating analytical components included into data pipeline over different clouds; however, provisioning resources from different clouds while meeting user performance requirements is also a challenge.} Combining these challenges and complexities with available heterogeneous compute resources that are available in cloud datacenters and users' quality of service requirements, managing the execution of such applications is a complex task. 

\par In general, stream workflows have received less attention, but the importance of making real-time decisions by analysing streaming data is rising with IoT emergence. Most of existing research works focused on supporting the other type of big data processing which is batch processing. These works such as \cite{wang2009kepler+} \cite{wang2014big} provided the ability to compose batch processing applications into pipeline to process static data at once and get final analytical insights by extending the capability of scientific workflow management systems. While the rest offered big data orchestrators (Apache YARN \cite{vavilapalli2013apache}, Apache Mesos \cite{hindman2011mesos} and Amazon Lambda \footnote{https://aws.amazon.com/lambda/details/}) that do not need to deal with the dynamism of stream workflow applications and meet real-time user requirements. Therefore, there are few scheduling algorithms in the literature that treat the scheduling problem of various streaming big data applications over cloud infrastructure.

Dealing with dynamic stream workflow application should be based on the fact that this workflow is an adaptive workflow application that serves the current-extra and future demands of changing real-time analytical requirements at runtime to make faster and better decisions.
\par To fill the gap of supporting dynamic scheduling under the variations of data stream rates, we design a new adaptive scheduling technique. This technique revises the scheduling plan of dynamic stream workflow application according to changes happen in the speed of data at runtime to always meet real-time analytical requirements with minimal execution cost. In other words, it is aimed at tackling data stream velocity fluctuations while maximizing performance efficiency, and all of that at minimal monetary cost. In summary, our contributions are: 
\begin{itemize}
	\item Dynamic stream application model.
	\item Two-phase adaptive scheduling technique that incorporates two advanced optimization algorithms (Random immigrants Genetic Algorithm (GA) and two-level Greedy algorithm) to efficiently execute dynamic stream workflows. 
\end{itemize}

\par This paper is structured as follows: Section \ref{DynamicSchedulingRelatedWork} reviews the related works. Section \ref{DynamicSchedulingWorkflowApp} presents dynamic stream workflow requirements. The problem formulation is presented in  Section \ref{DynamicSchedulingProbelmModeling}. Section \ref{DynamicSchedulingProposedTechnique} presents the proposed scheduling technique whose performance is evaluated in Section \ref{DynamicSchedulingExperimentsandDiscussion}. Section \ref{DynamicSchedulingConclusion} concludes the paper and highlights future improvements.

\section{Related Work} \label{DynamicSchedulingRelatedWork}

\textcolor{black}{In this section, we present the comparisons with related works from three perspectives, which are application, modelling and methods/techniques.}

\textcolor{black}{From application perspective, there are batch-oriented big data workflow (MapReduce workflow) and stream-oriented big data workflow (stream workflow). The focus of previous studies (such as J. Wang et al. \cite{wang2009kepler+}  \cite{wang2014big}, F. Teng \cite{teng2013scheduling}, Y. Wang and W. Shi \cite{wang2014budget}, T. Shu and C.Q. Wu \cite{shu2017performance}, and X. Zeng et al. \cite{zeng2016sla} \cite{zeng2018cost}) were mostly on MapReduce workflows and their executions in cloud computing infrastructure.}

\par \textcolor{black}{From modelling perspective, there are two stream processing models, which are data-flow graph with micro-batch processing model (i.e. discretized streaming model) and operator graph with continuous processing model. With discretized streaming model, streaming computations are performed on a series of small data batches called micro-batches. M. Zaharia et al. \cite{zaharia2012discretized} followed this model and proposed a stream programming model named Discretized Streams (D-Streams). It brings together a series of Resilient Distributed Datasets (RDDs) and allows performing computations through various transformations. Apache Spark uses RDD data model and allows to perform stream computations on RDD to define data processing. While with continuous processing model, operator graph is used to model data pipeline, where each node in the graph is a long-lived operator. This operator carries-out stream computation on streams as they arrive and produces new stream. Stream-oriented big data platforms and services such as Apache Storm and IBM Streaming allow to build streaming operator graphs for performing real-time data processing. As streaming operator graphs are different from dynamic stream workflows in that the source of data for the whole operator graph is one and there is one end operator, a new model is needed for dynamic stream workflow, which involves heterogeneity, multiple data sources and outputs.}

\par \textcolor{black}{From scheduling perspective, scheduling methods and techniques in the literature use heuristic and/or meta-heuristic approaches for making decisions based on different scheduling criteria (such as deadline, execution cost and performance) in order to meet user-defined SLA requirements. Research works such as D. Sun \cite{sun2015re}, T. Buddhika et al. \cite{buddhika2017online} and A. Bożek and F Werner \cite{bozek2018flexible} focused on scheduling data stream computations for performance and/or energy optimizations. Also, stream-oriented big data platforms and services such as Apache Storm and IBM Streaming allow to build streaming operator graphs to process data streams and produce final output stream. However, those research works and frameworks model stream workflow as a streaming operator graph. Since streaming operator graphs are different from dynamic stream workflows, the placement problem (i.e. scheduling problem) of dynamic stream workflows have different assumption and optimization goals. This problem considers the mapping of analytical components to multiple compute resources as well as the optimization goals include minimizing execution cost and improving performance without violating real-time user requirements.}

\par \textcolor{black}{In the same perspective, D. Sun and R. Huang \cite{sun2016stable} and D. Sun et al. \cite{sun2018rethinking} focused on online scheduling with guaranteed makespan and utilized single cloud as an execution environment for big data streaming application. These scheduling strategies/methods do not consider stream workflow as a network of streaming big data workflow applications (i.e. workflow of workflows). They also do not take into consideration the dynamic nature of this workflow and its unpredictable performance, the various real-time decision support requirements and the powerful capability of 'cloud of clouds' as a dynamic execution environment. In the same context but for scheduling big data processing jobs/tasks and workflow in geo-distributed clouds, L. Chen et al. \cite{chen2018scheduling} proposed fair job scheduler with the aim of reducing job completion time that relied on Apache Spark. Z. Hu et al. \cite{hu2016flutter} proposed a new job scheduling method named Flutter, which aimed at reducing completion time and implemented in Apache Spark. H. Chen et al. \cite{chen2018big} proposed task-duplication based real-time scheduling method to reduce completion and execution times. However, these scheduling methods are considered stream workflow as operator graph and have different optimization goals.}

\par \textcolor{black}{For scheduling techniques supported with big data application orchestrators, } each one of them uses a different scheduling technique to map applications on cloud resources. Apache YARN uses a monolithic scheduler to map compute resources among competing applications in the cluster. Apache Mesos uses a dual-level scheduling mechanism called “resource efforts”. This mechanism provides resource offerings to a framework and lets this framework either to accept the offer or reject it if the offered resources do not meet its constraints and then waiting for the ones they do. Consequently, these orchestrators assume either that they do not need to meet real-time decision support requirements or are intended for big data workflows that have predictable performance \cite{ranjan2017orchestrating}. Therefore, the scheduling mechanisms in those orchestrators consider big data workflow application as a static structure, so that they neglect the following: (1) dynamic nature of this application and its analytical components, (2) unpredictable performance of this workflow application, (3) real-time performance requirements defined by the owners of these workflows, (4) runtime changes and (5) the powerful capability of 'cloud of clouds' as a dynamic execution environment.

\textcolor{black}{Accordingly, the scheduling techniques proposed in the aforementioned studies do not fit the composition needs of complex big data workflows. They also do not leverage the capability of Multicloud environment to cope with the dynamic aspects of these workflows. As a result, dynamic scheduling technique is needed for a stable and efficient execution of stream workflow over multiple cloud infrastructures that meets user real-time performance requirements and respond to the runtime changes in velocity of streaming data while reducing the overall execution cost.}

\section{Dynamic Stream Workflow Application and its Requirement} \label{DynamicSchedulingWorkflowApp}

\textcolor{black}{Stream workflow application is a network of streaming big data applications (analytical components) that can be independently executed over cloud resources while maintaining data dependencies among them. It has three main characteristics that need be considered, which are continuous input data (from external and internal sources), continuous processing and continuous insights produced by end analytical components. Considering these characteristics, this application is a dynamic workflow, as the most dynamism form that occurs frequently is changing the velocity of streaming data for analytical components. The real use case for stream workflow application along with details of the requirements are provided in Appendix A. 
}

\textcolor{black}{In this workflow, the throughputs of services under the dynamic variations of input data rates should be maintained all the time. Moreover to achieving these throughputs, end-to-end latency (response time) is crucial in stream workflow application. It is a time between receiving input data stream at a service and generating output stream that regards this stream. Ensuring low latency is required during the whole execution of stream workflow. It should be kept as low as possible or be bounded when it starts to increase whilst maintaining user specific throughput.} \textcolor{black}{ Furthermore in this workflow, there are distributed data sources that inject their data streams into a data pipeline, thus data locality approach should be utilized by leveraging Multicloud environment. With this environment, we can avoid transferring large data to the corresponding resources over long-heal networks that is not only incurring high latency and execution cost, but also makes achieving real-time data analysis requirements more difficult. Additionally, each cloud in Multicloud environment offers different computing capabilities with different prices, so that changing the placement cloud for a service from one cloud to another is possible to reduce execution cost.}

\par Consequently, the variables of dynamic stream workflow application are type of service, its data processing requirement, its data processing rate, data mode, the dynamic variations of input data rates and the dynamism of execution environment (i.e. Multicloud). As well, variables of the later involved bandwidth and latency among various cloud infrastructures. Overall, the requirements of both workflow application and real-time data analysis should be maintained while dealing with data stream velocity fluctuations and minimizing the monetary cost during the continuous execution of this application.

\section{Problem Modelling} \label{DynamicSchedulingProbelmModeling}

\par Prior to introduce the problem modelling of stream workflow application, we list all the terminologies that will be used in this model in Table \ref{tab:notation}.

\begin{table}
	\scriptsize
	\centering
	\begin{center}
		\caption{\textcolor{black}{Problem Modelling Notation}}
		\begin{tabular}{|p{1.7cm}|p{6.3cm}|}
			\hline 
			Symbol / Term & Description \\ 
			\hline 
			G & Workflow graph \\
			\hline 
			S & Set of all graph services \\
			\hline 
			E & Set of all graph edges \\
			\hline 
			$\textyen^m$ & Percentage of data that is routed from parent service to child service (100\% in replica mode or any percent in partition mode) \\
			\hline 
			$S_n$ & Particular service in workflow graph \\
			\hline 
			$MI^{S_n}$ & Number of floating-point operations required to process one MB of input data (MI/MB)\\
			\hline 
			$\lambda^{S_n}$ & Amount of data produced by a given external source and being consumed by a service $S_n$ (MB/s) \\
			\hline 
			$\gamma^{S_n}$ & Proportion of output data to input data for $S_n$\\
			\hline 
			C & Set of all clouds in Multicloud environment \\
			\hline 
			$c_g$ & Particular Cloud in Multicloud environment\\
			\hline 
			L & Network latency matrix \\
			\hline 
			B & Network bandwidth matrix \\
			\hline 
			D & Data transfer cost matrix \\
			\hline 
			$VM^g$ & Set of all VMs in cloud g \\
			\hline 
			$vm_k^g$ & Particular VM k in cloud g \\
			\hline 
			$U^g$ & Set of all internal network links between VMs in cloud g \\
			\hline 
			$u_h^g$ & Particular internal link between $vm_{org}^g \text{ and } vm_{dest}^g$ \\
			\hline
			$MIPS_{vm_k^g}$ & Rating of the capacity of VM k in cloud g \\
			\hline 
			$\cent_{vm_k^g}$ & Provisioning cost of VM k in cloud g (cents/s) \\
			\hline 
			minDPUnit & Minimum stream unit for the whole application (MB) \\
			\hline 
			unitDPRate & Minimum stream processing rate based on minDPUnit for the whole application (MB/s) \\
			\hline 
			$ \vartheta^{S_n}$ & P minDPUnits based on percentage change from original data rate that being increased or decreased from input stream of service $S_n$ \\
			\hline
		\end{tabular}
		\label{tab:notation}
	\end{center}
\end{table}

\subsection{Application Model}

\textcolor{black}{Stream workflow application can be represented as a Direct Acyclic Graph (DAG) with \(G = (S, EX,E)\). S represents a set of N services \(S = {s_1, s_2, ..., s_N}\), EX represents a set of P external sources \(EX = {ex_1, ex_2, ..., ex_P}\) and E represents a set of M edges/links between external sources and services, and between services themselves \(E = {e_1,e_2, ..., e_M}\). Each edge, \(e_m\) is represented as a tuple \((\psi^m, s_{dest}^m, \textyen^m)\), where \(\psi^m\) denotes stream output source which is either \(ex^m\) denotes external source or \(s_{org}^m\) denote origin service, \(s_{dest}^m\) denotes destination service and \(\textyen^m\) denotes the percentage of data generated by \(\psi^m\) that is routed towards \(s_{dest}^m\).  \\}

\textcolor{black}{Each particular external source \(EX_p\) is represented as a tuple \(EX_p=(\Lambda^{EX_p}) \), where \(\Lambda^{EX_p}\) denotes the output data rate (data velocity) of this output source.} Each particular service \(S_n\), is represented as a tuple \(S_n = (MI^{S_n}, \lambda^{S_n}, \gamma^{S_n})\), where \(MI^{S_n}\) denotes the number of floating-point operations required to process one MB of incoming data (service data processing requirement) in MI/MB, \(\lambda^{S_n}\) denotes the arrival rate of data streams generated by sources outside the application in MB/s (such as data streams generated by sensors) to be consumed by the service, and \(\gamma^{S_n}\) denotes the proportion of data generated by the service based on input streams.\\

Notice that, given the nature of stream workflow applications, it is possible that data generated by one service can be sent to one or more services, or can be split among different services. Thus, for service \(S_n\), both parameters \(\gamma^{S_n}\) and \(\textyen^m\) (in edges where such service is origin service) are necessary to define the whole application. In addition, to process streams that coming at different speeds, the minimum stream unit per second (denoted as unitDPRate) is needed to be specified for the whole workflow application. Thus, each provisioned compute resource must process at least one unit per second, and of course, it can process multiple units per second according to its computing capacity.

\subsection{System Model}
The cloud system is modelled as a tuple W = (C, L, B, D). A set of G clouds in the Multicloud environment is denoted as \(C = {c_1, c_2, ..., c_G}\). L, B, and D denotes matrices containing respectively the latency (in seconds), the bandwidth (in MB/s), and the data transfer cost (in cents/MB or \cent/MB) between each of the pair of clouds in C.

Each cloud, \(c_g\) is represented as a tuple \( (VM^g, U^g) \), where \( VM^g = {vm_1^g, vm_2^g, ..., vm_K^g} \) is a set of K virtual machines (compute resources) with different resource configurations deployed in \(c_g\), and \( U^g = {u_1^g, u_2^g, ..., u_H^g}, u_h^g = (vm_{org}^g, vm_{dest}^g) \), a set of H links that are part of the data center network topology.

Each VM deployed in the cloud, \(vm_k^g\), is represented as a tuple \( (MIPS_{vm_k^g}, \cent_{vm_k^g})\), where \(MIPS_{vm_k^g}\) denotes floating-point operations computed by this VM according to its compute capacity per second and \(\cent_{vm_k^g}\) denotes the cost of provisioning such VM (in cents per second).

The data processing rate for \(S_n\) if it is mapped to \(vm_k^g\) is denoted as \(\varphi_k^g\) and is calculated \textcolor{black}{by dividing VM computing power by service unit data processing rate and service data processing requirement as follows:}

\begin{equation} 
\varphi(S_n, vm_k^g) = \frac{\lfloor MIPS_{vm_k^g} / \chi \rfloor \ast \chi}{MI^{S_n}} \text{  MB/s}
\end{equation}
\[Where \text{     } \chi = unitDPRate \ast MI^{S_n} 
and \text{     } MIPS_{vm_k^g} \geq \chi
\]

As \(S_n\) could be mapped to more than one VM to achieve user performance requirements, let \(pro(S_n)\) be the set of VMs that are provisioned from one cloud for service \(S_n\). The data processing rate for \(S_n\) if it is mapped to VMs in \(pro(S_n)\) \textcolor{black}{is the total data processing rates for \(S_n\) on all provisioned VMs and is calculated as follows:}
\begin{equation} \label{eq:mapping}
\varphi(S_n, pro(S_n)) = \sum\nolimits_{v \in pro(S_n)} \varphi(S_n, v)
\end{equation}
\textcolor{black}{In dynamic stream workflow application, the calculation of data processing rate for each service \(S_n\) should be carried-out at runtime. This is because of the need to handle dynamic changes that result in varying the speed of input streams being injecting into this service. Thus, system should calculate this rate based on the updated input speed of a service after the occurrence of change request at runtime. Let \(inStream(S_n)\) denote the input stream of \(S_n\) and is the total rate of incoming data from external sources and internal sources (i.e. parent services) based on data modes used to route such streams toward this service:}

\begin{equation}
\begin{multlined}
\textcolor{black}{inStream(S_n) = \lambda^{S_n} + \mathop{\sum\nolimits_{e_x \in E | \psi^x=s_{org}^x \& s_{dest}^x=S_n}}} \\ \textcolor{black}{{(\gamma^{S_{org}^x} \ast \varphi(S_{org}^x, pro(S_{org}^x))) \ast \textyen^{S_{org}^x} \textrm{ MB/s}}} \\
\textcolor{black}{\textit{Where } \lambda^{S_n} = \mathop{\sum\nolimits_{e_x \in E | \psi^x=ex^x \& s_{dest}^x=S_n}} (\Lambda^{ex^x} )}
\end{multlined}
\end{equation}

The following data processing constraint of \(S_n\) is maintained:
\begin{equation} \label{eq:mappingConstraint}
\varphi(S_n, pro(S_n)) \geq inStream(S_n)
\end{equation}
Each service \(S_n\) produces output stream as a result of computation. Let \(outStream(S_n)\) denote the output data stream for a service \(S_n\) and is calculated \textcolor{black}{by multiplying the total input rate of \(S_n\) by output data proportion/percent as follows:}
\begin{equation} \label{eq:5}
outStream(S_n) = \gamma^{S_n} \ast inStream(S_n) 
\end{equation}

The velocity of data for given external source may change at runtime, which leads to a direct impact either an increase or decrease on the velocity of data for each \(S_n\) connected to this source. This change makes \(inStream(S_n)\) and \(outStream(S_n)\) be updated by the amount of data that being increased or decreased. Also, this change affects not only those services, but also has a subsequent change (i.e indirect impact) on the velocity of data for child services which have dependency-link with those services. Therefore, it is worth to note that the maximum number of velocity changes that can be sent at any instant of time is assumed to be one and such velocity change request (either increase or decrease request) is only happen via external source. Let \(\vartheta^{S_n}\) denote the amount of data stream (in MB/s) based on percentage change from original data rate that being increased or decreased to \(inStream(S_n)\) as P minDPUnits. In case of data velocity decrease, $\vartheta^{S_n}$ should be $0 < \vartheta^{S_n} < inStream(S_n)$.
\textcolor{black}{The \(inStream(S_n)\) will be updated by adding or subtracting P minDPUnits and \(outStream(S_n)\) will be updated by multiplying the update total input rate of \(S_n\) by output data proportion/percent as follows:}
\begin{equation}
\begin{array}{l}
inStream(S_n) = inStream(S_n) \pm \vartheta^{S_n}\\ 
outStream(S_n) = \gamma^{S_n} \ast inStream(S_n)
\end{array}
\end{equation}

As well as the the decrease in velocity of data for \(S_n\) leads to lower computing needs for maintaining the above data processing constraint, so that VM(s) that are not required will be deprovisioned. This results in cost reduction while meeting user real-time data processing requirements. While the increase in velocity of data leads to more computing demands to maintain the above data processing constraint for this high data rate, so that exVM(s) will be provisioned. Let \(exVM(S_n)\) be the set of new VMs that need to be provisioned from placement cloud of service \(S_n\) to cope with the increase speed of data streams, and \(rmVM(S_n)\) be the set of VM(s) from \(pro(S_n)\) for service \(S_n\) that will be terminated/deprovisioned in response to an decrease in the speed of data streams. Thus, pro(\(S_n\)) is updated periodically at runtime \textcolor{black}{by provisioning new VM(s) in case of velocity increases or deprovisioning VM(s) from the existing ones to respond to velocity decreases as follows: }

\begin{equation}
pro(S_n) = 
\begin{cases}
pro(S_n) + exVM(S_n),  \textit{if velocity incr.} \\
pro(S_n) - rmVM(S_n), \textit{if velocity decr.}\\
pro(S_n), \textit{otherwise (no change)}
\end{cases} 
\end{equation}

As pro(\(S_n\)) is updated at runtime, the \(\varphi(S_n)\) is also updated, 
reflecting the new data processing rate for \(S_n\) based on the updated pro(\(S_n\)). 

Given the change in velocity of data that either increases or decreases data rate which leads to provision more VMs or deprovision existing VMs at runtime, the execution cost needs to be calculated frequently. For our problem here, the calculation basis for the total execution cost of dynamic stream workflow application is per second. If T is total time duration, for cost calculations it is divided into several one second intervals (i.e. \(T_1, T_2, ..., T_i\)).

\textcolor{black}{Additionally, we assume that every data stream should be processed, as unprocessed data streams lead to incorrect results. We also assume that the order of stream portions should be maintained during the distributed among the corresponding compute resources. Based on these assumptions, we maintain user specific throughputs for all services and end-to-end latency (response time) as low as possible or even bounded when it is being increased. Thus, the incoming data streams are processed as they arrive and the latency is maintained, which is a time from a stream being added to input queue until its emission from the service as output stream. Of course, in case of a child service receives two or more dependency streams from its parents services, the latency is from the time of the last stream being added to input queue until its emission from child service.}

The cost of running VMs used by service \(S_n\) to process incoming streams per second \(T_i\) is denoted as ec(\(S_n\)) while the total cost of running all VMs used by all services to process incoming streams during period of time T is denoted as ExecCost(S,T). \textcolor{black}{The ExecCost(S,T) is calculated by summing VM provisioning costs for all services for T time as follows:}

\begin{equation}
\begin{multlined}
ExecCost(S,T) = \sum\nolimits_{T_i} \sum\nolimits_{S_n} ec(S_n) \qquad \textrm{cents}
\end{multlined}
\end{equation}

\textcolor{black}{The ec(\(S_n\)) is calculated by totalling the costs of all VMs provisioned for \(S_n\) per second as follows:}
\begin{equation}
\begin{multlined}
ec(S_n) = \sum\nolimits_{v \in pro(S_n)} \cent_{v} \qquad \textrm{cents}
\end{multlined}
\end{equation}

The data transfer cost is based on the amount of data being moved, the cost of data transfer charged by cloud provider, and network bandwidth. In a dynamic workflow application, the velocity of data determines the speed of generation, processing and analysis of data, where both input and output data are moved among different clouds. As we mentioned before, the change in velocity of data affects the data transfer cost as increasing speed leads to an increase in the cost and vice versa, so that the cost calculation needs to be carried-out per second. Let \(cts(S_n)\) denote the cost of transferring streams for \(S_n\) (including input streams from other services) per second, and \(CTStream(S, T)\) denote the total data transfer cost for the amount of data being moved for all services during the period of time T. \textcolor{black}{The \(CTStream(S,T)\) is calculated by summing the costs of data transfer between services for T time as follows:}

\begin{equation}
CTStream(S, T) = \sum\nolimits_{T_i} \sum\nolimits_{S_n} cts(S_n) \qquad \textrm{cents} 
\end{equation}

\textcolor{black}{The cts(\(S_n\)) is calculated by totalling the costs of data transfer performed by \(S_n\) per second as follows::}
\begin{equation}
cts(Sn) = \sum\nolimits_{S_i \in parent(S_n)} c(S_i)  \textrm{    cents} 
\end{equation}

\begin{equation*}
c(S_i) =  
\begin{cases}
0,& \text{if } C_g(S_i) = C_g(S_n)\\
outStream'(S_i) \\ \ast D(C_g(S_i),\\C_g(S_n)), & \text{otherwise}
\end{cases}
\end{equation*}

\[
outStream'(S_i)
\begin{cases}
outStream(S_i),& \text{if } \varrho \leq 1\\

\frac{outStream(S_i) \ast \textyen^x}{\varrho} , & \text{otherwise}
\end{cases}
\]

\[ \textit{Where } \varrho = \frac{outStream(S_i)\ast \textyen^x}{B(C_g(S_i), C_g(S_n))} + L(C_g(S_i), C_g(S_n))\] 
\[ \textit{      , and } parent(S_n) \textit{is the set of parent services for service } S_n\] 

\textcolor{black}{Overall, the objective function is to minimize the cost of executing the dynamic workflow without violating data dependences and real-time performance requirements while dealing with changes in speed of data at runtime:}
\begin{equation} \label{eq:objective}
\begin{multlined}
min f(S,T) = ExecCost(S,T) + CTStream(S, T)
\end{multlined}
\end{equation}

\par \textcolor{black}{Eq. \ref{eq:objective} is solved for minimization to generate cost-efficient scheduling plan for the execution of dynamic stream workflows. Considering services' data processing requirements and the variety of resources offered by multiple clouds, each service can be mapped to more than one resource in order to maintain its data processing constraint based on input data rate (refer to Eq. \ref{eq:mapping} and Eq. \ref{eq:mappingConstraint}). If we relax such mapping constraint thus each service is mapped only to one resource (i.e. $\vert pro(S_n)\vert = 1$), assuming that this resource is sufficient to meet service's data processing constraint (Eq. \ref{eq:mappingConstraint}), this relaxed constraint makes the problem 0-1 assignment problem. In this problem, the assignment matrix M indicates that a service $i$ is assigned to resource $j$ if $m_{ij} = 1$. This problem is well-known NP-hard \cite{martello1981algorithm}. Consequently, if we consider the mapping of a service to more than one resource (i.e. not relaxing mapping constraint) now, our problem is even harder than 0-1 assignment problem, so it is NP-hard problem. Moreover, our problem belongs to NP because if a feasible resource allocation solution is given, this solution can be tested in polynomial time using Algorithm \ref{alg:polyAlgo}. Accordingly, our problem is NP-complete problem.}

\begin{algorithm}
	\footnotesize
	\begin{algorithmic}[1]

		\STATE totalDPRate $\gets$ 0
		\FOR{each service $S_n$ in S}
			\FOR{each VM $vm_k^g$ from $prov(S_n)$}
				\STATE totalDPRate = totalDPRate + $\varphi(S_n, vm_k^g)$ 
			\ENDFOR
			\IF{totalDPRate $< inStream(S_n)$}
				\STATE return false \COMMENT{this is not  feasible solution}
			\ENDIF
		\ENDFOR				
	\end{algorithmic}
	\caption{\textcolor{black}{polynomial-time algorithm for checking the feasible solution}}
	\label{alg:polyAlgo}
\end{algorithm}

\section{Proposed Adaptive Scheduling Technique} \label{DynamicSchedulingProposedTechnique}

\par \textcolor{black}{As we discussed in the previous section, our scheduling problem is NP-complete problem. Thus, the problem's search spaces are complex, with large sets of VM offerings provided by various cloud infrastructures and many constraints that need to be fulfilled such as data dependencies, user-defined real-time performance, throughput and end-to-end latency. Indeed, the search space of finding candidate solution for efficient execution of dynamic stream workflow application rapidly increases with the size of problem. Furthermore, the fluctuation of data velocity overtime makes it necessary to re-explore the complex search space in order to find sub-optimal solution as quick as possible, where exhaustive search for the optimal solution is not feasible. Consequently, th goal is to find near-optimal solution in the complex search space and revise it as fast as possible to tackle the changes in data velocity overtime  without violating data dependences and real-time performance requirements while minimizing the total execution cost (Eq \ref{eq:objective}).} As we cope with velocity change for this workflow application, the following are cases of changing in input stream rate of a service: 

\begin{itemize}
	\item The speed of output stream of external source connected to this service is either increased or decreased.
	\item The speed of output stream of parent service(s) connected to this service is either increased or decreased. This happen when the increase or decrease in the speed of stream propagated from parent services due to the increase or decrease in the speed of stream for connected external sources
\end{itemize}

\par \textcolor{black}{From the aforementioned goal, we have two challenges: (1) explore large search space to find candidate solution at deployment time and (2) revise this solution quickly and precisely with each velocity change request that occurs at runtime to locate sub-optimal solution to respond to such request. For the first challenge, genetic algorithm is useful algorithm in exploring complex search space to enable the practical implementation of optimizing problem; thus, the objective function of Eq. \ref{eq:objective} can be considered as a fitness function of genetic algorithm.  While for the second challenge, Greedy heuristic can be used to adopt deployment plan generated by genetic algorithm at runtime because it provides an immediate sub-optimal solution for tackling velocity change request as it needs a relatively small time to compute; thus, it can fulfil the need to make scheduling decision under time constraints, enabling the practical implementation of optimizing objective function at a given point. Accordingly, we propose a new adaptive scheduling technique for dynamic stream workflows.}

\par
\begin{figure}
	\centering
	\includegraphics[width=1\linewidth]{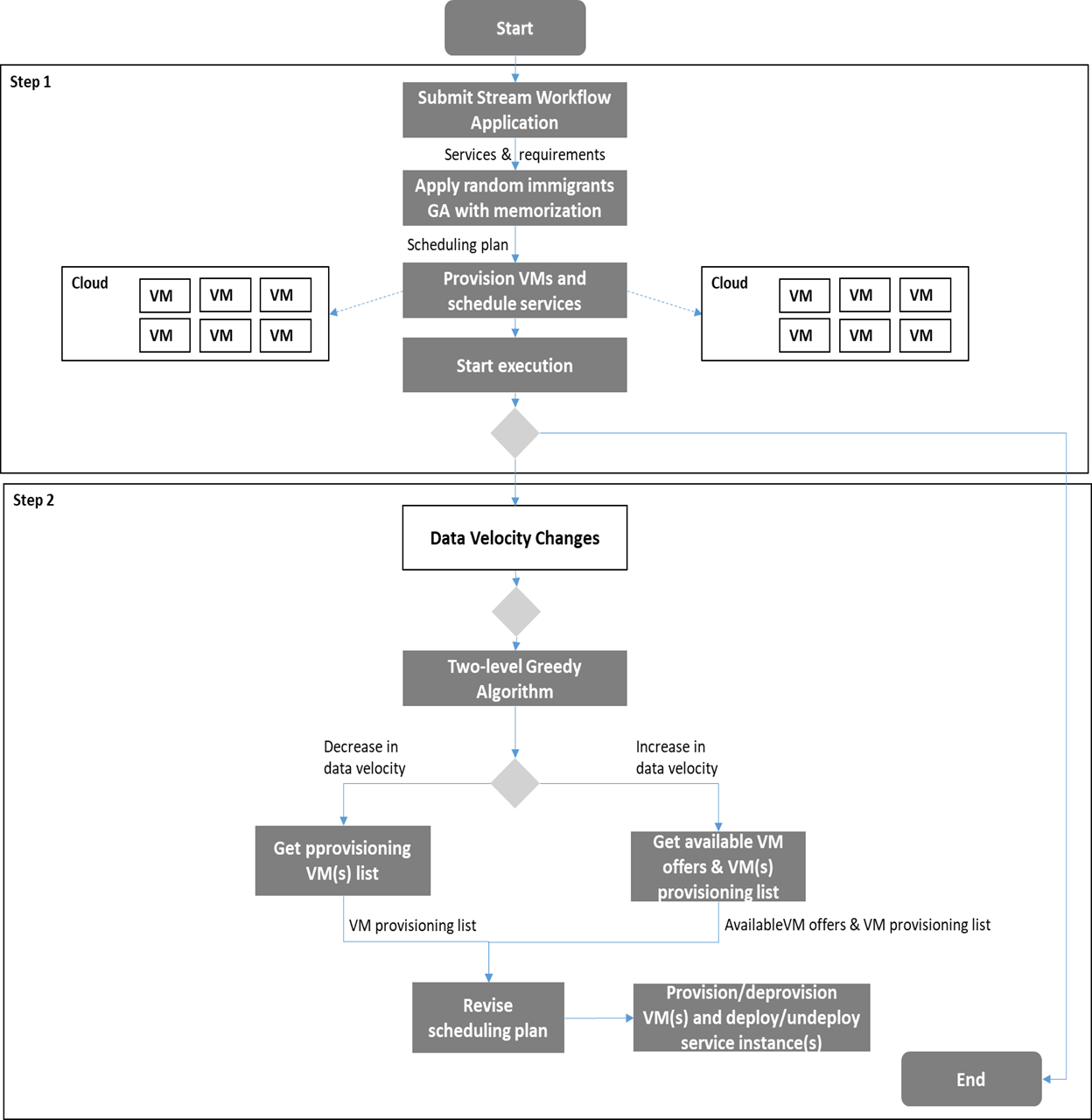}
	\caption{\textcolor{black}{Proposed two-phase scheduling technique flowchart}}
	\label{fig:twostepschedulingflowchart}
\end{figure}

 The proposed technique is a two-phase dynamic workflow scheduling technique that incorporates two advanced optimization algorithms (i.e. Random immigrants Genetic Algorithm (GA) in Phase 1 and two-level Greedy algorithm in Phase 2) to effectively perform adaptive scheduling of dynamic stream workflow applications in Multicloud environment and intelligently response to changes happen at runtime (i.e. velocity changes) with minimal execution cost. \textcolor{black}{The main design goal of this technique is to find the best placement plan for the services of given workflow application with minimal execution and data transfer costs and maintaining its efficiency after each adaptation to handle the velocity change requests. The flowchart of the proposed two-step scheduling technique is depicted in Figure \ref{fig:twostepschedulingflowchart}. In the below paragraphs, we will discuss the two steps (and their sub-steps) of this technique.} that exploiting the deployment flexibility provided by Multicloud environment.	

\textcolor{black}{\textit{First Step} $-$ The proposed random immigrants GA is called to find the best global sub-optimal resource selection solution according to the original real-time performance requirements to improve scheduling efficiency. Once the stream workflow application is scheduled on provisioned resources and being executed, the proposed technique is moved to the next step to tackle dynamic changes in the velocity of data streams.}

\textcolor{black}{\textit{Second Step} $-$ In this step, the proposed two-level Greedy algorithm is used to dynamically respond to the changes in the speed of data streams for services. This algorithm at first level determines the services whose their input data rates will be changed due to the received velocity change request. Then in second level, it finds the best resource provisioning/deprovisioning solution(s) that will be used to tackle these changes and revising scheduling plan (to provision new VMs or deprovision existing ones that are not needed any more).}

\subsection{GA with Random Immigrants Scheme}

\par Traditional GA has a considerable problem, which is convergence that prevents genetic diversity of the population. Therefore, to avoid such problem and to enhance the genetic diversity of the population, random immigrants schema is used \cite{yang2008genetic}. This schema retains diversity level of the population every generation via replacing a portion of candidate solutions in the current population with random candidate solutions called immigrants. Accordingly, we propose a random-based immigrants GA (GA for short) that is able to find sub-optimal resource selection solution for scheduling dynamic stream workflow application in Multicloud environment. It exploits data locality by selecting the most appropriate datacenter for each service, which leads to the reduction in both execution and data transfer costs. It generates the initial population randomly, and then evaluate the candidates and sort them in acceding order of fitness. During each generation, the elite candidate is selected and m random immigrants are generated then replaced the worst n candidates in the current population. Following the evaluation of m immigrants, the selection, crossover and mutation operators are applied. Finally, the elite candidate is added and the evolved population is evaluated and then sorted in acceding order of fitness before go to the next generation. \textcolor{black}{The time complexity of this algorithm is presented in Table \ref{tab:GATimeComplexity}.} The Watchmaker framework for evolutionary computation \cite{watchmaker2010framework} is used to implement this algorithm.

\begin{table}[H]
	\scriptsize
	\centering
	\caption{\textcolor{black}{Time complexity of random-based immigrants GA}}
	\begin{tabular}{|p{20em}|p{10em}|}

		\hline
		Name  & Time complexity \\
		\hline
		Random population generation & $O(su)$ \\
		\hline
		Fitness Function & $O(ps^2d)$ \\
		\hline
		Roulette wheel selection with binary search & $O(plog(p))$ \\
		\hline
		Crossover & $O(s)$ \\
		\hline
		Mutation & $O(sv)$ \\
		\hline
		Sort & $O(plog(p))$ \\
		\hline
		Random-based immigrants schema & $O(s^2d)$ \\
		\hline
		\textbf{Total}  & $O(gps^2d)$ \\
		\hline
		\multicolumn{2}{|p{30em}|}{\textit{g} the number of generations (as termination condition), \textit{p} the size of population, \textit{s} the length of candidate solution (number of services), \textit{u} is maximum number of required minimum data processing units among services, \textit{v} is number of VM offers in the placement cloud and \textit{d} is maximum number of stream dependencies among services} \\
		\hline
	\end{tabular}%
	\label{tab:GATimeComplexity}%
\end{table}%

\subsection{Two-level Greedy Algorithm}

\par We propose a new two-level greedy algorithm that uses Minimax with Alpha-Beta pruning method in game theory to minimize the maximum resource provisioning cost by finding the best resource selection solution for services that affected by data velocity changes. Minimax with Alpha-Beta pruning method is considered as a powerful searching and decision-making algorithm on game tree to find optimal/sub-optimal result from possible choices. Thus, this method is used in our algorithm to find the best resources with the lowest provisioning cost at runtime to achieve the updated data processing rate for each service affected directly and indirectly by velocity change request. The direct effect happens when the service is connected to external source whose data velocity will be changed, while indirect effect occurs when the service is in the velocity change path). 

Our proposed algorithm addresses the problem of ongoing resource scaling under the dynamic variations of data stream rates by managing resources overtime. This algorithm at first level determines the services whose their input data rates will be changed due to the received velocity change request. Then, at second level, it finds the best resource provisioning/deprovisioning solution(s) that will be used to revise scheduling plan. With the occurrence of velocity change request, it finds the best provisioning and scheduling solution, and then dynamically and quickly updating the scheduling plan to respond to this change request while reducing the overall provisioning cost. \textcolor{black}{The pseudocode of the proposed two-level greedy algorithm is shown in Algorithm \ref{alg:twolevelGreedy} \textcolor{black}{ and the time complexity analysis of this algorithm is presented in Table \ref{tab:TwoLevelGreedyTimeComplexity}.} The pseudocode of two procedures that used in this algorithm to respond to velocity increase and decrease requests are shown in Algorithm \ref{alg:increaseVelocityProc} and Algorithm \ref{alg:descreaseVelocityProc} respectively. The pseudocode of Minimax with Alpha and Beta algorithm that used in both procedures (Algorithm \ref{alg:increaseVelocityProc} and Algorithm \ref{alg:descreaseVelocityProc}) is shown in Algorithm \ref{alog:MinimaxAlphBetaAlgorithm}. Algorithm \ref{alog:MinimaxAlphBetaAlgorithmEvaluateFunction} shows the pseudocode of evaluation function used in Algorithm \ref{alog:MinimaxAlphBetaAlgorithm}.}

\begin{algorithm}
	\footnotesize
	\begin{algorithmic}[1]
		\STATE $min \gets TreeNode(-1, \infty)$ \COMMENT{-1 is vm global id and $\infty$ is value}
		\STATE $max \gets TreeNode(-1, -\infty)$ \COMMENT{-1 is vm global id and -$\infty$ is value}
		\STATE $depth \gets 2$ \COMMENT{depth level in game tree}
		\STATE $ \textit{affectedSIDs} \gets \text{get ids of services affected by velocity change request}$ \label{findAffectedServices}
		\FOR{each service $S_n$ in affectedSIDs }

		\IF{velocity change request is increase request} \label{increaseRequest}
			\STATE Velocity\_Increase\_Req\_Proc($S_n$, min, max, depth)
		
		\ELSE \label{decreaseRequest}
			\STATE Velocity\_Decrease\_Req\_Proc($S_n$, min, max, depth)
		
		\ENDIF
		\ENDFOR

	\end{algorithmic}
	\caption{\textcolor{black}{Two-level Greedy Algorithm}}
	\label{alg:twolevelGreedy}
\end{algorithm}

\begin{algorithm}
	\footnotesize
	\begin{algorithmic}[1]

		\STATE $ reqUnits \gets 0 $
		\STATE $ unitMIPS \gets MI^{S_n} * unitDPRate $
		\STATE $ avalVms \gets \text{get VM offers from service placement cloud} $
		\STATE $ avalVms = avalVms - \{x \in VM^g | MIPS_x < unitMIPS\} $
		\STATE $ extraAchievedUnits \gets \varphi(S_n, pro(S_n)) / minDRRate - \lceil (inStream(S_n) * MI^{S_n}) / unitMIPS \rceil $

		\STATE $ incRate \gets \text{get data rate increases over service input rate} $

		\STATE $ reqUnits \gets \text{get number of units required based on incRate} $
		\STATE $ reqUnits \gets reqUnits - extraAchievedUnits $
		\STATE $ nodes \gets \text{create tree nodes for avalVms list} $
		\WHILE{$ reqUnits > 0 $}
		\STATE $ \text{shuffle nodes and construct tree with specified depth} $

		\STATE $ root \gets \text{get root of constrcuted tree}$
		\STATE $ best \gets \text{Minimax\_AlphaBeta(depth, true, root, min, max)} $ \COMMENT{best node for VM selected}
		\STATE $ VMList = VMList \cup best.getVmgid() $
		\STATE $ reqUnits = reqUnits - \lfloor (MIPS_{best.getVm()} /unitMIPS) \rfloor $
		\ENDWHILE

		\STATE $\text{add VMList of } S_n \text{ to ServiceVMsMap}$ \COMMENT{ VMList$\neq \phi$}
	\end{algorithmic}
	\caption{\textcolor{black}{Velocity\_Increase\_Req\_Proc($S_n$,min,max,depth)}}
	\label{alg:increaseVelocityProc}
\end{algorithm}

\begin{algorithm}[t!]
	\footnotesize
	\begin{algorithmic}[1]
		
		\STATE $ redUnits \gets 0 $
		\STATE $ unitMIPS \gets MI^{S_n} * unitDPRate $
		
		\STATE $ \text{SPVMs} \gets pro(S_n) $

		\STATE $ extraAchievedUnits \gets \varphi(S_n, pro(S_n)) / minDRRate - \lceil (inStream(S_n) * MI^{S_n}) / unitMIPS \rceil $
		\STATE $ decRate \gets \text{get data rate decreases from service input rate} $
		
		\STATE $ redUnits \gets \text{get number of minDPUnits based on decRate} $
		\STATE $ redUnits \gets redUnits + extraAchievedUnits $
		\WHILE{$ redUnits > 0 $}
		\STATE $ \text{remove VM(s) from SPVMs achieved } units>redUnits$
		
		\IF{ \text{SPVMs is empty}}
		\STATE return \COMMENT{no provisioned VM can be deprovisioned}
		\ENDIF
		\STATE $ \text{construct tree nodes from SPVMs list with specified depth} $
		
		\STATE $ root \gets \text{the root of constrcuted tree}$
		\STATE $ best \gets \text{Minimax\_AlphaBeta(depth, true, root, min, max)} $ 
		\STATE $ VMList = VMList \cup best.getVmgid() $
		\STATE $ redUnits = redUnits - \lfloor (MIPS_{best.getVm()} /unitMIPS) \rfloor $
		\STATE $\text{SPVMs = SPVMs } - \text{ best.getVm()}$
		\ENDWHILE
		\IF{VMList is not empty}
		\STATE $\text{add VMList of } S_n \text{ to ServiceVMsMap}$
		\ENDIF
	\end{algorithmic}
	\caption{\textcolor{black}{Velocity\_Decrease\_Req\_Proc($S_n$,min,max,depth)}}
	\label{alg:descreaseVelocityProc}
\end{algorithm}

\begin{algorithm}[t!]
	\scriptsize
	\caption{Minimax\_AlphaBeta(depth, maximizingPlayer, node, alpha, beta)}
	\begin{algorithmic}[1]
		\IF{depth == 0}
		\STATE $\textit{return evaluate(node)}$
		
		\ELSIF{maximizingPlayer}
		\FOR{each child of node }
		\STATE $\textit{ TreeNode val = } \newline \textit{Minimax\_AlphaBeta(depth - 1, false, child, alpha, beta)}$		
		\IF {\textit{val.getValue() $>$ alpha.getValue()}}
		\STATE $\textit{alpha = val}$	
		\ENDIF				
		
		\IF {\textit{beta.getValue() $<=$ alpha.getValue()}}
		\STATE $\textit{break}$	\COMMENT{alpha cut-off}
		\ENDIF
		\ENDFOR
		\STATE $\textit{return alpha}$
		
		\ELSE
		\FOR{each child of node}
		\STATE $\textit{ TreeNode val = } \newline \textit{Minimax\_AlphaBeta(depth - 1, true, child, alpha, beta)}$		
		\IF {\textit{val.getValue() < beta.getValue()}}
		\STATE $\textit{beta = val}$	
		\ENDIF				
		
		\IF {\textit{beta.getValue() <= alpha.getValue()}}
		\STATE $\textit{break}$	\COMMENT{beta cut-off}
		\ENDIF
		\ENDFOR
		\STATE $\textit{return beta}$
		\ENDIF

	\end{algorithmic}
	\label{alog:MinimaxAlphBetaAlgorithm}
\end{algorithm}

\begin{algorithm}[t!]
	\scriptsize
	\caption{Evaluation Function - evaluate(node)}
	\begin{algorithmic}[1]
		\REQUIRE
		\STATE $\textit{reqUnits, redUnits, unitMIPS}$
		\STATE $value, cost \gets 0$ \COMMENT{value for increase request and cost for decrease request}	
		\IF {\textit{velocity change request is increase request}}
		\STATE $\textit{VMboottime } \gets \text{get boottime for VM node}$	
		\STATE $\textit{achievedUnits } \gets \text{get units achieved by VM node}$	
		
		\STATE $\textit{value} \gets (achievedUnits / (reqUnits \ast \cent_{vm_k^g})) / \textit{VMboottime} $
		\STATE $\textit{value} \gets value +  \lfloor MIPS_{vm_k^g} / (unitMIPS \ast \#OfServiceDependencies) \rfloor / \cent_{vm_k^g} $
		\STATE $ node.value \gets value$	
		\ELSE
		\STATE $\textit{achievedUnits } \gets \text{get units achieved by VM node}$	
		\STATE $\textit{cost} \gets (achievedUnits / (redUnits \ast \cent_{vm_k^g})) $
		\STATE $ node.value \gets cost$	
		\ENDIF
		\STATE $ \textit{return node}$
	\end{algorithmic}
	\label{alog:MinimaxAlphBetaAlgorithmEvaluateFunction}
\end{algorithm}

\par Prior processing the velocity change request, the proposed technique finds the ids of service affected by this request directly or indirectly (Algorithm \ref{alg:twolevelGreedy} Line \ref{findAffectedServices}). Then, for each service affected, it finds the best provisioning or deprovisioning solution based on the type of velocity change request. If the request is velocity increase request (Algorithm \ref{alg:twolevelGreedy} Line \ref{increaseRequest}), \textcolor{black}{it calls Algorithm \ref{alg:increaseVelocityProc} to get} VM offers of service placement cloud and then finding the extra minDPUnits that are achieved by the current provisioned VMs in accordance to service input data rate. Next, such algorithm calculates the number of minDPUnits required for data rate being increased over service input data and then deducting from this number the extra achieved units. After that, it calls Algorithm \ref{alog:MinimaxAlphBetaAlgorithm} several times to finds best VM(s) to provision until achieving the required units. While, with velocity decrease request (Algorithm \ref{alg:twolevelGreedy} Line \ref{decreaseRequest}), \textcolor{black}{it calls Algorithm \ref{alg:descreaseVelocityProc} to get} the list of VMs provisioned for a service and then finding the extra minDPUnits that are achieved by these VMs based on service input data rate. Next, such algorithm calculates the number of minDPUnits based on the data rate being decreased from service input data, and then increasing this number by extra achieved units. Next, it removes those VMs from the list of provisioned VMs where their powers achieved units greater than the number of minDPunits that will be removed. The remaining VMs in this list will be used to find the best VM to deprovision using  \ref{alog:MinimaxAlphBetaAlgorithm}. 

\par Each run of the game finds the best VM to provision it in case of velocity increases or to deprovision it in case of velocity decreases. Since multiple VMs may be needed to achieve the updated data processing rate or may be released in response of decreasing the velocity, the game will be repeated to produce the best solution. For each VM selected, the number of minimum data processing units achieved based on the computing power of this VM in one game is deducted from the total required units (i.e. reqUnits) in case of velocity increases or from total reduced units (i.e. redUnits) in case of velocity decreases.

\begin{table}[H]
	\scriptsize
	\centering
	\caption{\textcolor{black}{Time complexity of two-level greedy algorithm}}
	\begin{tabular}{|p{20em}|p{10em}|}

		\hline
		Name  & Time complexity \\
		\hline
		Get affected services & $O(s)$ \\
		\hline
		Velocity increase request procedure & $O(ub^m)$ \\
		\hline
		Velocity decrease request procedure & $O(ub^m)$ \\
		\hline
		Minimax alpha-beta & $O(b^m)$ \\
		\hline
		Evaluation function & $O(1)$ \\
		\hline
		\textbf{Total} & $O(sub^m)$ \\
		\hline
		\multicolumn{2}{|p{30em}|}{\textit{s} number of services, \textit{u} is maximum number of required minimum data processing units among services, \textit{b} is branching factor and \textit{m} is maximum depth of the tree} \\ 
		\hline
	\end{tabular}%
	\label{tab:TwoLevelGreedyTimeComplexity}%
\end{table}%

\section{Experiments and Discussion} \label{DynamicSchedulingExperimentsandDiscussion}

\subsection{Experiment Methodology}

\subsubsection{Configuration of Workflow Application}

The four well-known workflow structures from various domains are Montage in Astronomy, Inspiral in Astrophysics, Epigenomics in Bioinformatics and Cybershake in Earthquake science. These workflows operate on static data inputs and produce outputs. To simulate stream workflow applications, it is possible to use these workflow structures to achieve that by considering each job as a service and the flow of data to be streams of data \textcolor{black}{coming from external and internal sources rather than static files}.

In addition, a set of extra parameter configurations is required for those workflow structures to simulate stream workflow applications. These parameter configurations include data processing requirements of services, data rates of external sources, types of services, modes of data, placement datacenter (in case of unmovable service) and output data rates of services. As a result, different stream workflow applications can be modelled using the aforementioned workflow structures (Montage, Inspiral, Epigenomics and CyberShake) by simulating them in these workflow applications for our experiments. For each workflow structure, three different sizes are used (small, medium and large). Therefore, 12 real workflow applications are modelled for our experiments, which are Montage\_25, Inspiral\_30, Epigenomics\_24, CyberShake\_30, Montage\_50, Inspiral\_50, Epigenomics\_46, CyberShake\_50, Montage\_100, Inspiral\_100, Epigenomics\_100, CyberShake\_100.

\subsubsection{Multicloud Environment}

To form a Multicloud environment for our experiments, we model three different cloud system providers, namely (Amazon EC2 \cite{Amazon2017Instances}, Google Cloud Engine \cite{Google2017Instances}, and Microsoft Azure \cite{Microsoft2017Instances}). Each cloud system has different VM configurations that chosen from pre-defined machine types offered by this cloud provider. \textcolor{black}{The details of computing power rating being used and different VM configurations offered by those clouds are provided in Appendix B.}

\par In addition, to model boot time (startup time) for each VM configuration in the modelled clouds, we use average range of VM startup time defined in \cite{ContainersOrVirtualMachines2015}. For each modelled cloud, we generate random numbers from the defined range and then assign these numbers to its VM configurations.

\subsubsection{Configuration of Data Velocity}

\par To model the amount of data that is being increasing or decreasing in velocity change request for one external source, we utilised future data rates given in Gartner foreseen \cite{hassan2017internet} which specifies one connected vehicle will generate as much as 25GB/hour of data, equivalent to 8MB/s. By considering this value as the average data rate of external source in workflow application, we create different percentage ranges for modelling the increase and decrease in data velocity. For velocity increase, we model the value of increase in data velocity as a percentage that is increased from current data rate. We model the data velocity decrease as percentage of decrease in the current data rate. Table \ref{tab:ConfigDataVelocityIncrease} lists the percentages of change to increase data velocity. Table \ref{tab:ConfigDataVelocityDecrease} shows change percentages  to decrease data velocity. It is worth to note that as there is a minimum limit for stream unit, the change value will be approximated/rounded to the nearest given minDPUnit. As an instance, if the minimum stream unit per second is 1MB/s and the 65\% increase in data velocity from 5MB/s as original data rate is chosen randomly, the approximation will be applied on the change value (3.25MB/s) to be 3MB/s (i.e. the nearest value based on the specified minDPUnit) so that the new data rate will be 8MB/s.

\begin{table}[H]
	\scriptsize
	\centering
	\caption{Percentage ranges of data velocity increase amount}

	\begin{tabular}{|c|c|c|}

		\hline
		Velocity Range & Minimum (Percent) & Maximum (Percent) \\
		\hline
		Low     & 10   & 30 \\
		\hline
		Medium  & 50   & 70 \\
		\hline
		High    & 90    & 100 \\
		\hline
	\end{tabular}%
	\label{tab:ConfigDataVelocityIncrease}%
\end{table}%

\begin{table}[H]
	\scriptsize
	\centering
	\caption{Percentage ranges of data velocity decrease amount}

	\begin{tabular}{|c|c|c|}

		\hline
		Velocity Range & Minimum (Percent) & Maximum (Percent) \\
		\hline
		Low     & 5   & 15 \\
		\hline
		Medium  & 25   & 35 \\
		\hline
		High    & 45    & 50 \\
		\hline
	\end{tabular}%
	\label{tab:ConfigDataVelocityDecrease}%
\end{table}%

\subsubsection{Workflow and Simulation Parameters}

\par To run our experiments, we need to configure a set of parameters for both workflow application and simulator. These parameters and their values are fixed for all scenarios and listed in Table \ref{tab:WorkflowAndSimulationParamters}. For external data source rate, the value considered is from the data velocity configuration discussed in the previous subsection. For network bandwidth and latency for ingress and egress traffic, we conducted TCP bandwidth and
latency tests between different zones of Nectar Cloud \footnote{https://nectar.org.au/research-cloud/} and considered the medium range from obtained results. For the cost of data transfer, we find the minimum and maximum data transfer costs between modelled clouds, and then use the created medium range to obtain values. Service data processing requirement represents the complexity of computation that carried-out by this service, which varies from simple (20 MI/MB) to complex (4000 MI/MB) aggregation functions. For the value of this requirement, we consider the created medium range.

\begin{table}
	\tiny
	\centering
	\caption{Workflow and simulation parameters}
	\begin{tabular}{|p{18em}|p{20em}|}

		\hline
		Parameter & Value \\
		\hline
		External Source Data Rate  & 5 MB/s with increase-velocity experiment \newline 10 MB/s with decrease-velocity experiment \\
		\hline
		Ingress Network Bandwidth  & Range [615, 926] MB/s \\
		\hline
		Ingress Network Latency  &  Range [0.00064, 0.00086] second \\
		\hline
		Egress Network Bandwidth  & Range [122, 218] MB/s \\
		\hline
		Egress Network Latency  & Range [0.021, 0.031] second \\
		\hline
		Data Transfer Cost & Ingress traffic: 0 \newline Egress traffic: Range [0.013 - 0.019] cents/MB \\
		\hline
		Type of Service & 50\% unmovable services 50\% movable services \\
		\hline
		Service Data Processing Requirement & Range [1348, 2674] MI/MB \\
		\hline
		Service Data Processing Rate & System-calculated rate based on input stream(s) \\
		\hline
		Data mode type & Replica \\
		\hline
		Service Output Data Rate &  Range [1, 50] \% of input rate \\
		\hline
		Minimum Data Processing Unit & 1 MB \\
		\hline			
		Minimum Data Processing Rate &  1 MB/s \\
		\hline
		GA - Population Size & 50 \\
		\hline
		GA - Generation Limit & 50 \\
		\hline
		GA - Elitism & 1 \\
		\hline
		GA - Crossover Probability  & 0.8 \\
		\hline
		GA - Mutation Probability  &  0.3 \\
		\hline
		GA - Number of Random Immigrants & 5 \\
		\hline
		Number of Velocity Change Events  & 2 \\
		\hline
		Delay between Velocity Change Events  & 10 seconds \\
		\hline
		Simulation Time  & 180 seconds (3 minutes) \\
		\hline
	\end{tabular}%
	\label{tab:WorkflowAndSimulationParamters}%
\end{table}%

\subsubsection{Experimental Scenarios}

\par Our experimental evaluations for efficiency and performance of the proposed technique are described in the below paragraphes:

Comparison with baseline, GA and lower bound (Evaluation 1) $-$ Study and compare the proposed adaptive scheduling technique (GA + Two-level greedy algorithm) in finding the best resource provisioning solution and adapting scheduling plan in response to velocity increases/decreases with competitors (baseline algorithm and random-based immigrants GA scheme) and lower bound. This comparison in term of the execution cost of different workflow applications for 3 minutes simulation time. \textcolor{black}{A realistic baseline algorithm is created for our problem that does not need to use any heuristic.} It finds VM with the highest computing power and then provisioning it to respond to velocity increase requests, while with velocity decrease requests, it deprovisions one or more VMs from the available VMs to respond to these requests. \textcolor{black}{The aim of comparison with baseline algorithm is to appreciate the necessity of our proposed technique to find the best resource provisioning solution and adapting the scheduling plan in response to velocity increases/decreases.} \textcolor{black}{The comparison with GA schema is aimed at evaluating the proposed technique with another meta-heuristic algorithm that is widely used in workflow scheduling research works in order to further proof its efficiency}. Furthermore, the comparison with lower bound is to show that the complicated heuristic is necessary to approach the lower bound as well as to evaluate how the proposed technique is far from lower bound. In lower bound, we have relaxed many constraints. The first constraint is service's datacenter placement constraint and the second one is VM provisioning constraint (by selecting the cheapest VM across all datacenter VM offers). The third constraint is data transfer cost (by using a lower cost value from the specified range). The last constraint is network bandwidth constraint (by using a lower bandwidth from the specified range which leads to reduction in data transfer cost by transferring less data). Then for each service, the cheapest VM from the placement cloud of this service is provisioned as many as is required to achieve the specified data processing rate. After that, the total execution cost (provisioning cost + data transfer cost) is calculated using Equation \ref{eq:objective} during the period of time T. In this comparison, we consider the results obtained from lower bound as the base values.
	
\textcolor{black}{Guaranteeing processing speed for execution time (Evaluation 2) $-$ Study the efficiency of the proposed adaptive scheduling technique in guaranteeing processing speed required with different workflow applications under different velocity changes. \textcolor{black}{This evaluation aims to show the data processing constraint is satisfying at all time with changing data velocity.} The baseline here to achieve real-time user-defined requirements and end-to-end execution time is that the computing power available should be sufficient to process all incoming data without data loss. In other words, the computing capacity should be grater than or equal the velocity of incoming data at runtime. Thus, the sufficient computing capacity should be always maintained while the velocity of data increases or decreases at runtime. The experimental results will be collected before the end of simulation due to at that time all velocity changes have been made and handled by the proposed technique. This ensures the efficiency of the proposed technique to adopt scheduling plan in respond to velocity change request at runtime while guaranteeing processing speed required to achieve end-to-end execution time.}
	
\begin{figure}
	\centering
	\includegraphics[width=0.7\linewidth]{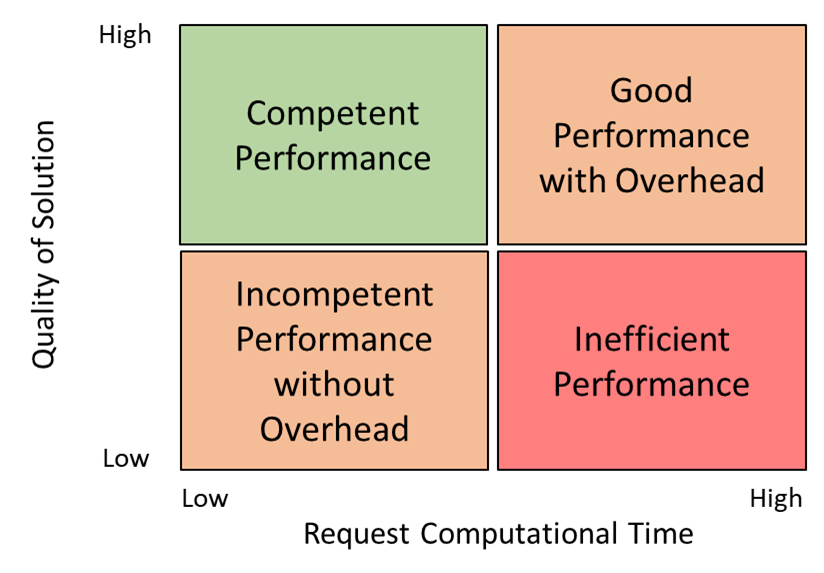}
	\caption{Performance Matrix}
	\label{fig:performancematrix}
\end{figure}

Efficiency of velocity change (Evaluation 3) $-$ Study and compare the proposed adaptive scheduling technique with random-based immigrants GA scheme (GA for short) based on performance matrix presented in Figure \ref{fig:performancematrix} for performing dynamic scheduling at runtime. \textcolor{black}{The aim of this evaluation is to determine how our proposed technique is effective in satisfying quality of service.} This comparison is in term of the quality of solution for the revised scheduling plan, which includes solution cost (provisioning + data transfer cost per second) after the data velocity change request is applied, and the number of changes applied on the current scheduling plan to respond to this change request. The GA responses to each velocity change request by generating a totally new scheduling plan, which serves as a revised plan to replace the old one, while the proposed technique revises the current scheduling plan.  \textcolor{black}{When GA applied those VMs in the new plan that exist in old plan are excluded to avoid VM duplication. By doing so,  only VMs in the new plan that are not exist in the old plan will be provisioned and those VMs in old plan that are not part of the new plan will be deprovisioned. Therefore, the number of changes include the changes in provisioning plan (for new VMs that are not in the old plan) and deprovisioning plan (for provisioned VMs that are not exist in the new plan).}

\par In aforementioned scenarios, data rate of each external source in workflow application is set to be 5MB/s with velocity-increase experiment or 10 MB/s with velocity-decrease experiment at the beginning of execution. As velocity change requests being sent, the data rate of chosen external sources will be increased or decreased according to the conducted experiment.

\par By comparing the total execution costs of all workflow applications obtained from proposed technique with the lower bound of total execution costs for these applications, we can evaluate the efficiency of the proposed technique in finding the best solution either resource provisioning or deprovisioning solution in response to velocity increases or decreases. In addition, comparing and evaluating the quality of solution for the proposed technique with GA allows to evaluate the performance of the proposed technique in relative to the performance of GA.

\subsection{Experimental Results}

\begin{figure*}
	\centering
	\includegraphics[width=0.8\linewidth]{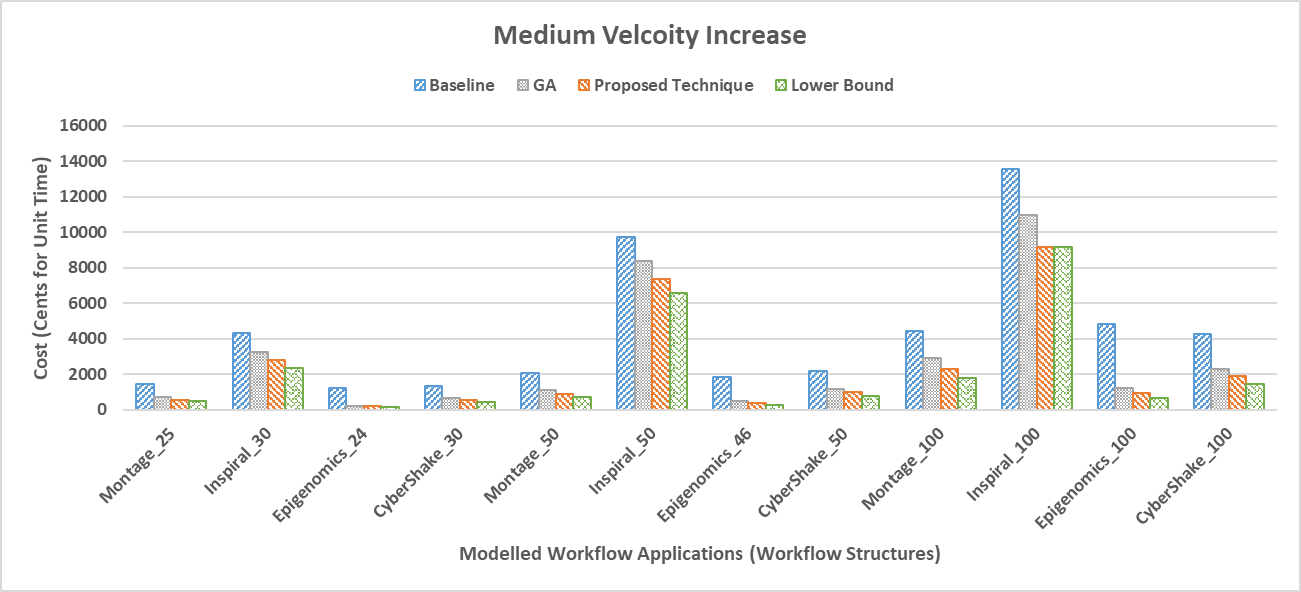}
	\caption{Total Execution Cost vs. Modelled workflow applications under medium range of velocity increase}
	\label{fig:Sceanrio_increase_medium}
\end{figure*}

\begin{figure*}
	\centering
	\includegraphics[width=0.8\linewidth]{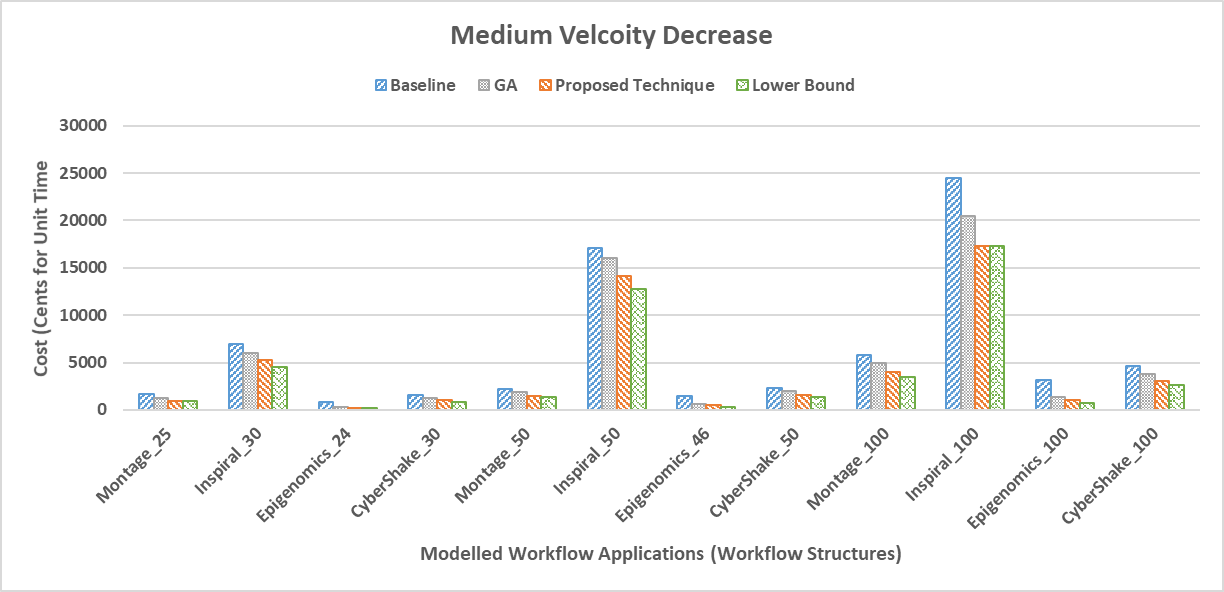}
	\caption{Total Execution Cost vs. Modelled workflow applications under medium range of velocity decrease}
	\label{fig:Sceanrio_decrease_medium}
\end{figure*}

\begin{figure*}
	\centering
	\includegraphics[width=0.8\linewidth]{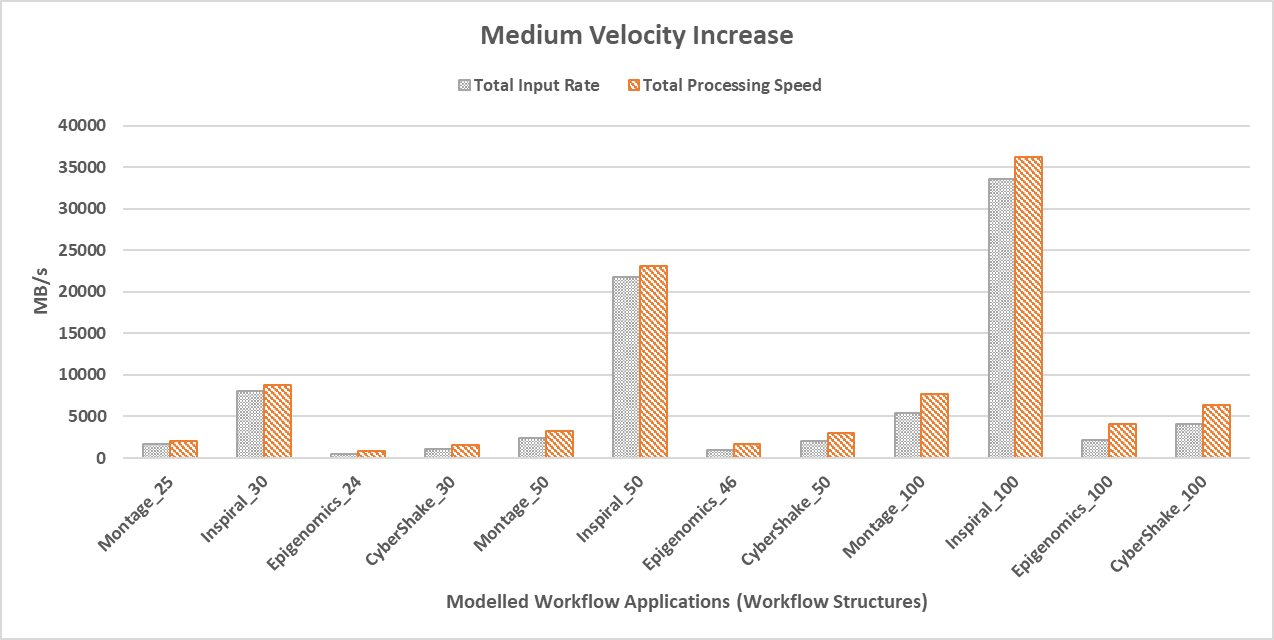}
	\caption{\textcolor{black}{Total Input Rate vs. Total Processing Speed for different workflow structures (medium velocity increase range)}}
	\label{fig:Sceanrio_Increase_TotalInputAndPP}
\end{figure*}

\begin{figure*}
	\centering
	\includegraphics[width=0.8\linewidth]{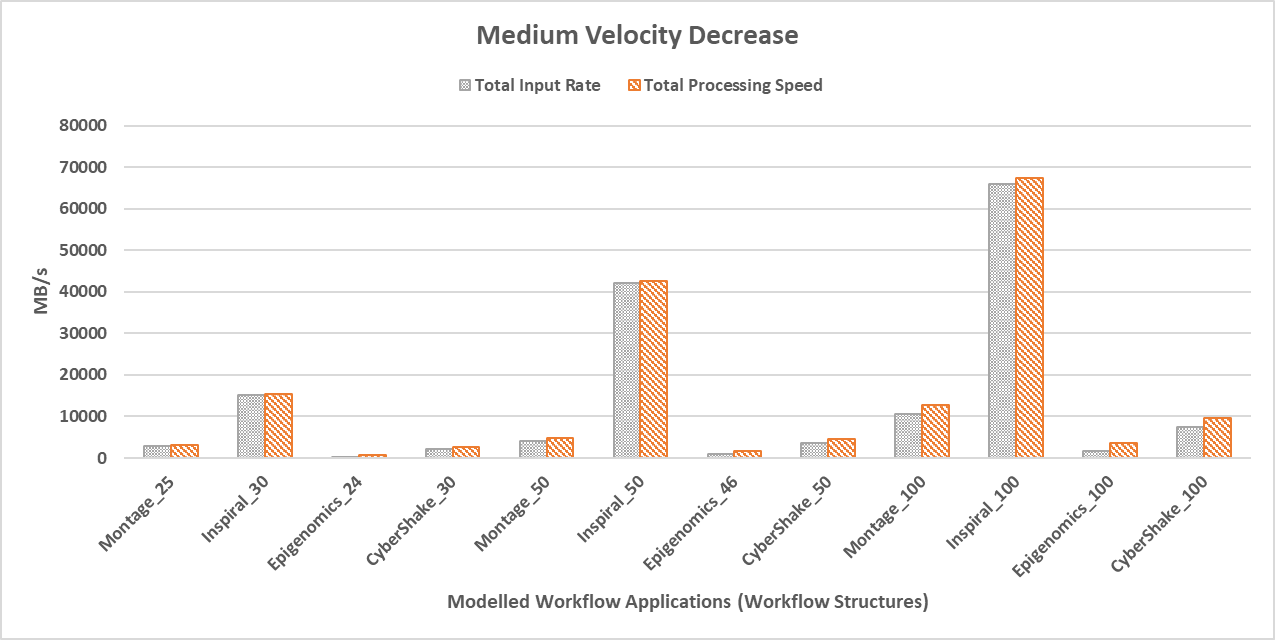}
	\caption{\textcolor{black}{Total Input Rate vs. Total Processing Speed for different workflow structures (medium velocity decrease range)}}
	\label{fig:Sceanrio_Decrease_TotalInputAndPP}
\end{figure*}

\par\textcolor{black}{To evaluate the efficiency and performance of the proposed technique, we conduct our experiments in simulation environment. This is because we need a controllable and repeatable environment to configure the parameters of each experiment scenario, and then compare the results obtained from the proposed technique with those from competitors under the same environment conditions. In real environment, some parameters like network bandwidth and latency cannot be controlled, making environment conditions are changing with each execution of workflow application. Thus, conducting our experiments in a real environment will produce inconsistent evaluation results, where these results cannot be used to assess the efficiency of proposed technique and the quality of solution produced to respond to data velocity change requests at runtime. Accordingly, we conduct our experiments using IoTSim-Stream \cite{barika2019iotsim}, our simulation toolkit for modelling and simulating stream workflow applications in Multicloud environments.}

The experimental scenarios are performed in simulation environment (by using IoTSim-Stream) on a Nectar Cloud virtual machine that had 8 vCPUs, 32GB of RAM memory and running Ubuntu 16.04.1 LTS, and the experimental results are collected. Since genetic algorithm is used in our proposed technique, each experimental scenario runs ten times, and the average value of the obtained results is taken and used in the representation of experimental results.

\subsubsection{Evaluation 1: Results}
\textcolor{black}{We conducted experiments to record total execution cost achieved by proposed technique and the competitors (Baseline, GA and Lower Bound) for modelled workflow applications under different ranges of velocity increase and decrease.} As the experimental results for the first evaluation showed that the total execution costs of modelled workflow structures under different velocity change ranges (low, medium and high) for both velocity-increase and velocity-decrease have not changed significantly, we only present those results for medium range of velocity change.

\par Figure \ref{fig:Sceanrio_increase_medium} and  \ref{fig:Sceanrio_decrease_medium} depict the total execution costs of modelled workflow applications under medium range of velocity increase and decrease that achieved by baseline algorithm, GA, proposed technique and lower bound. From these results, our analysis and findings are:

\begin{itemize}
	\item With various workflow applications, the proposed technique is efficient in finding the best solution to quickly respond to velocity change requests and then dynamically updating the current scheduling plan. The results of total execution cost obtained by the proposed technique compete the results obtained from both baseline and GA, and are close to the results of lower bound with most workflow structures. \textcolor{black}{The reason behind that is the proposed technique uses GA at first phase for exploiting data locality to find near-optimal placement and scheduling plan, which reduces resource provisioning and data transfer costs, and then in the second phase, it uses greedy heuristic to find the best provisioning plan that reduces the provisioning cost as much as possible to respond to any velocity change request.
	\item The cost resulting form the proposed technique is a maximum of 32\% of the cost generated by lower bound under medium velocity change. The reason for this difference is due to the structure of workflow may lead to process less data, so that the provisioning cost reduction factor contributes more to the total execution cost. Based on that, lower bound produces unachievable results as VM provisioning constraint is relaxed, while the proposed technique maintains this constraint.}
	\item \textcolor{black}{As data velocity increases from low to high range, the total execution cost for modelled workflow applications is slightly increased. The reason behind that is the proposed technique is able to revise the current plan to cope with velocity increase changes with minimal cost, leading to cost reduction even with high velocity of data.}  
	\item \textcolor{black}{The proposed technique is an adequate and practical dynamic scheduling method with competent accuracy. This is because it takes all the defined constraints into consideration while meeting user real-time performance requirements and reducing the overall execution cost with different workflows.}
\end{itemize}

\begin{figure*}
	\begin{subfigure}{1\textwidth}
		\centering
		\includegraphics[width=0.7\linewidth]{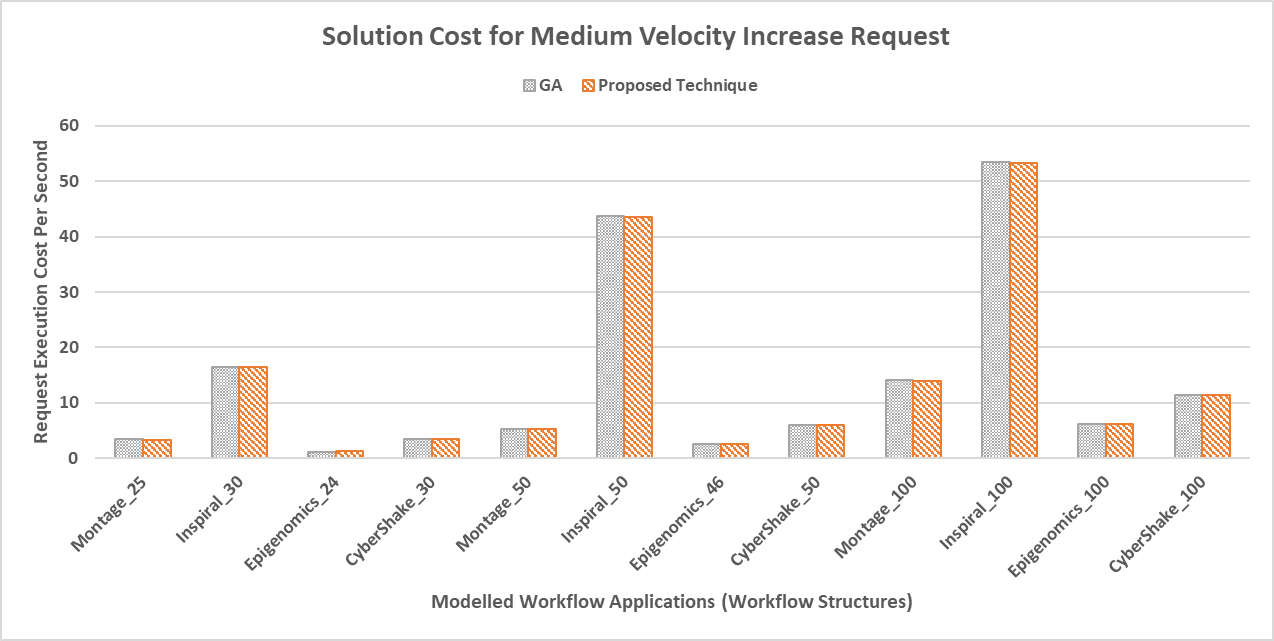}
		\caption{Solution cost}
		\label{fig:increaseECost}
	\end{subfigure}
	
	\begin{subfigure}{1\textwidth}
		\centering
		\includegraphics[width=0.7\linewidth]{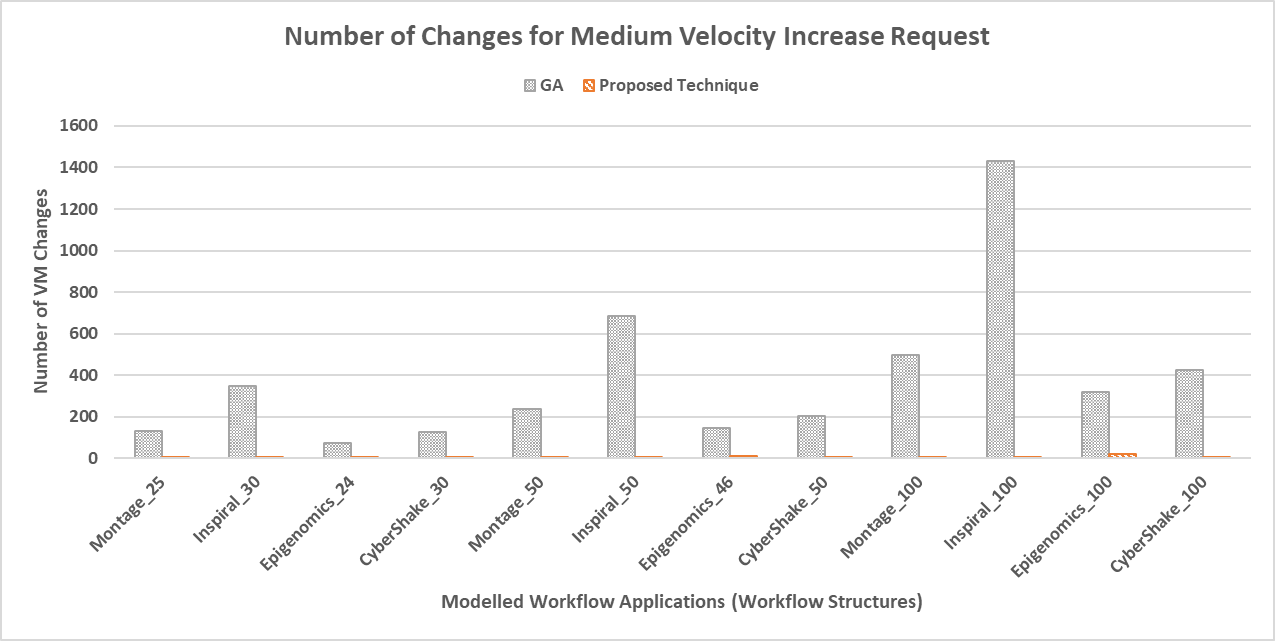}
		\caption{Number of changes}
		\label{fig:increaseNumbOfChanges}
	\end{subfigure}
	\caption{Quality of solution for different workflow structures (medium velocity increase range)}
	\label{fig:SecondEvaluation3_Increase}
\end{figure*}

\begin{figure*}
	\begin{subfigure}{1\textwidth}
		\centering
		\includegraphics[width=0.7\linewidth]{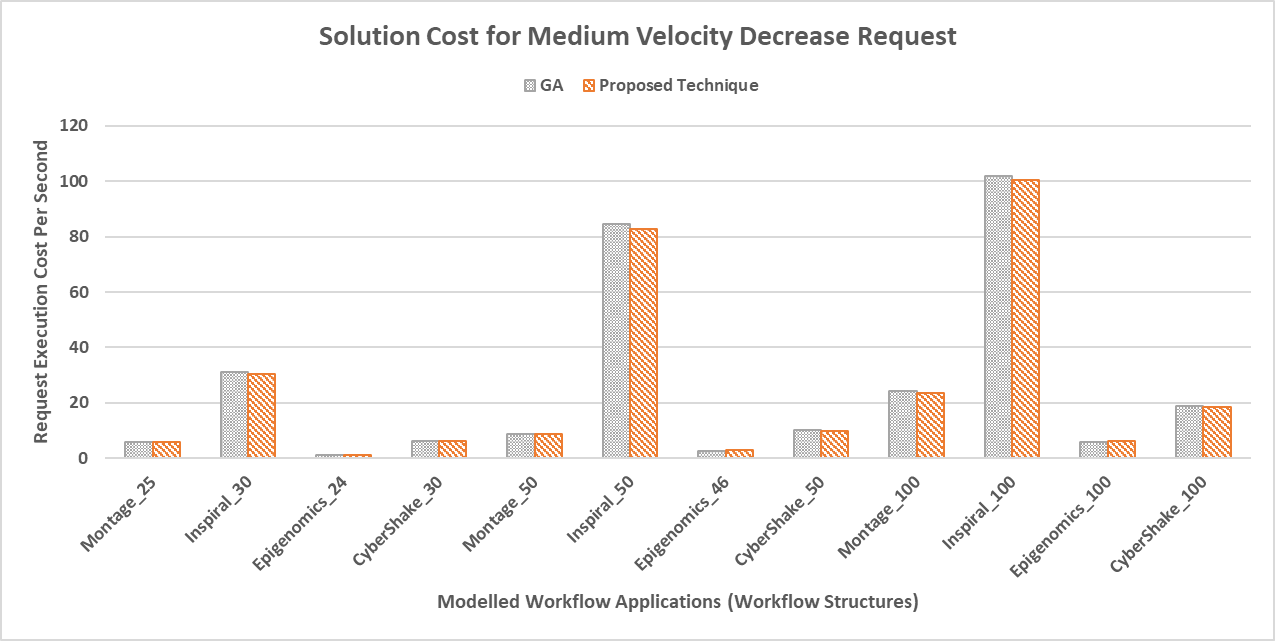}
		\caption{Solution cost}
		\label{fig:decreaseECost}
	\end{subfigure}
	\begin{subfigure}{1\textwidth}
		\centering
		\includegraphics[width=0.7\linewidth]{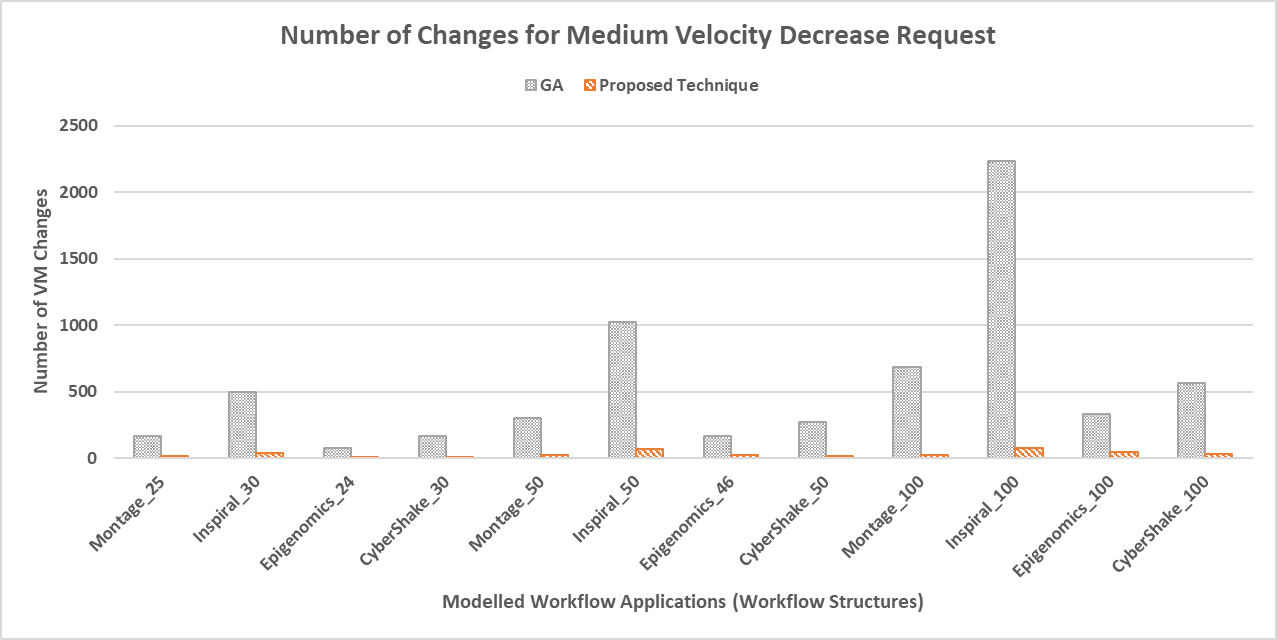}
		\caption{Number of changes}
		\label{fig:decreaseNumbOfChanges}
	\end{subfigure}
	\caption{Quality of solution for different workflow structures (medium velocity decrease range)}
	\label{fig:SecondEvaluation3_Decrease}
\end{figure*}

\subsubsection{Evaluation 2: Results}
\textcolor{black}{We conducted experiments to record total input data rate (in MB/s) and total processing speed (in MB/s) achieved by proposed technique for modelled workflow applications under different ranges of velocity increase and decrease. From the results obtained, we present here those results for medium range of velocity increase and decrease since these results are enough to reach to the conclusion. Figure \ref{fig:Sceanrio_Increase_TotalInputAndPP} and Figure \ref{fig:Sceanrio_Decrease_TotalInputAndPP} show the experimental results achieved by the proposed technique in term of total input rate and total processing speed. From the presented results, it is clear that the proposed technique always guarantee processing speed to process incoming data with all workflow application. Even more, it also has some extra computing power to handle increase in data velocity with immediate response and without the need to reschedule the execution plan.}

\subsubsection{Evaluation 3: Results}
We conducted experiments to collect solution cost and number of changes achieved by proposed technique and GA for modelled workflow applications under different ranges of velocity increase and decrease. It is worth to note that we do not need to conduct experiments to record end-to-end latency, since our assumption in problem modelling is that every data stream arrives will be processed as soon as the dependency is achieved. Moreover, we do not need to conduct experiments to collect execution time required to respond to velocity change request because of the conclusion is straightforward. In regards to the computational time for velocity increase request, the straightforward conclusion is that GA needs more time to generate a new scheduling plan to repose to this request, while the proposed technique fastly revises the current scheduling plan by replying on the two-level greedy algorithm for dynamic scheduling. In the favor of the velocity decrease request, the computational time needed for proposed technique to respond to this request is negligible since it just needs to deprovision the unnecessary VMs while GA needs to generate a new scheduling plan, so that it is far-fetched for GA to compete in that.

We present here the key important results from the second evaluation experiments. The experimental results for medium range of velocity changes (including both velocity increase and decrease requests) are only provided due to space reason. These results are enough to reach to the conclusion. Figure \ref{fig:SecondEvaluation3_Increase} and Figure \ref{fig:SecondEvaluation3_Decrease} show the experimental results achieved by the proposed technique in comparison to GA in term of solution cost and number of changes required to revise the current scheduling plan. From the presented results, the following are our analysis and findings:

\begin{itemize}
	\item In term of execution cost, GA with each request tries to find sub-optimal solution by generating a new plan whilst the proposed technique just revises the current plan quickly, which may not lead to sub-optimal solution. 
	\item In term of the number of changes for a velocity request, the proposed technique responds to this request by quickly adjusting the current scheduling plan instead of generating a completely new scheduling plan, which is usually required a limited number of VM changes. In contrast, GA generates a new scheduling plan in both velocity changes, which is not only incurring more computational time but also requires a lot of VM changes to deprovision those VMs that are not in the new plan and to provision those that are in the new plan. To maintain the continuity of processing incoming streams at current data rates, unneeded VMs from the old plan must remain in use until the new VMs become ready. This causes further overhead in execution time and cost more as both new VMs and the current VMs (that will be deprovisioned later on) are remaining in resource pool. Moreover, this also incurs more processing delays for upcoming streams when velocity change is increase request or more provisioning cost when the change is decrease request.
	\item \textcolor{black}{From Figure \ref{fig:increaseNumbOfChanges} and Figure \ref{fig:decreaseNumbOfChanges}, we can notice that the most performance gains is achieved by Inspiral\_100. The reason behind that is the structure of this workflow processes huge amount of data compared with other workflows, resulting more computing powers are required. Thus, generating a new plan is too expensive and incurring large number of changes, while revising the existing plan incurs small number of changes that leads to huge performance gains.}
	\item Based on the presented performance matrix (Figure \ref{fig:performancematrix}), the proposed technique achieved a competent performance with high quality of solution besides good execution cost compared to GA with most workflows, non-competitive number of changes required to revise the scheduling plan, and little or negligible execution time. Thus, the proposed technique outforms GA with different workflow structures.
\end{itemize}

\section{Conclusion and Future Work} \label{DynamicSchedulingConclusion}

In this paper, we considered the scheduling problem of dynamic stream workflow application on various Cloud infrastructures. These infrastructures forming a Multicloud environment, which becomes the dynamic execution environment for these applications. To this end, we proposed a new dynamic scheduling and provisioning technique that incorporates GA and two-level greedy algorithm to efficiently schedule dynamic stream workflow application in Multicloud environment while meeting real time user performance constraints under velocity changes with minimal execution cost. The experimental results showed that the proposed technique outperformed competitors (baseline and genetic algorithms) in responding to data velocity changes at runtime while reducing the total execution cost for all modelled workflow applications under various data velocity ranges. It also close to from lower bound.

For future study, this paper reveals two new directions to enhance the performance and capability of the proposed dynamic scheduling technique. The first direction is aiming to parallelise Minimax with Alpha-Beta pruning algorithm to reduce running cost and achieve speedup. The second direction is to support more dynamism forms for stream workflow applications such as application structure and real-time data processing requirement, where coping with these changes at runtime enables the full dynamic support.

\section*{Acknowledgment}
This research is supported by an Australian Government Research Training Program (RTP) Scholarship.

\bibliographystyle{IEEEtran}
\bibliography{references}

\appendices

\section*{Appendix}

\subsection{Real Use Case for Dynamic Stream Workflow}

\par
\begin{figure*}
	\centering
	\includegraphics[width=0.7\linewidth]{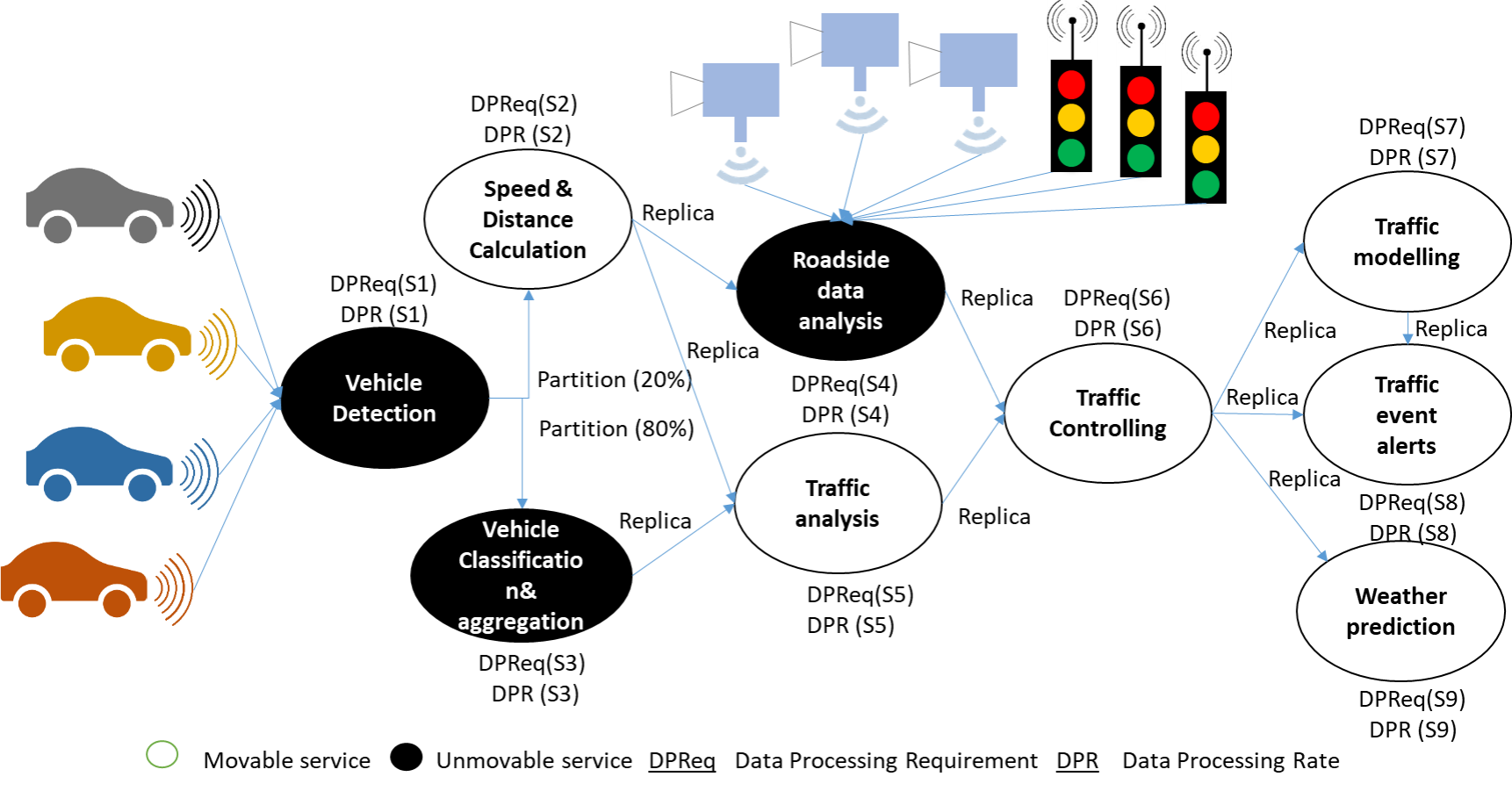}
	\caption{Dynamic stream workflow application example}
	\label{fig:streamworkflow}
\end{figure*}

 A real use case for big data streaming workflow application that shows the need of real-time analytics and workflow orchestration in the next era of technologies, consider connected vehicles in smart cities. Since the traffic is strained with the continued increase of the number of vehicles and population, the smart road traffic monitoring as a service of smart city services can utilize the true power of IoT connected vehicles in addition to roadside infrastructure (e.g. traffic lights, cameras). Collecting and analysing the streaming data generated by these vehicles allow to create real-time view of road traffic and incidents. Figure \ref{fig:streamworkflow} depicts streaming data pipeline for real-time view of road traffic in smart city. This pipeline represents a dynamic stream workflow application, which is a network of streaming analytical components. Each analytical component can be seen as a service because it can independently execute over any virtual resources, even though data dependencies among services should be maintained. 

\par In the presented workflow application, streaming data generated by connected vehicles on the road as sensor data is injected into vehicle detection service to detect the presence of vehicles in real-time. This service produces continuous output streams that are partitioned between two services for applying specific calculation, and classifying and aggregating vehicle data. The information of movement flow based on vehicle speed and distance as a result of continuous computations is further analysed by roadside data analysis and traffic analysis services. Based on this movement flow and sensor data coming from traffic lights and cameras, the roadside data analysis service produces real-time information about traffic density, which gives the opportunity for adjusting traffic patterns. Also, the traffic analysis service generates in-depth real-time analytics for traffic patterns and conditions by processing and analysing two data stream inputs, the movement flow information and the aggregated vehicle data. These analytics as output streams are injected into traffic controlling service for processing to improve traffic modelling, alert road users and authorities about traffic events and predict the current weather. In this application, the most dynamic form that occurs frequently is changing the velocity of streaming data for services. This is because the smart city is dynamic environment and the speed of streaming data is changing greatly based on time or traffic alert. Thus, the load at given time determines the required resources for computing. Thus, dynamic scaling and elastic of road traffic monitoring service application should be treated carefully to achieve real-time performance requirements under varying of data rates.

\par A service in the presented workflow has data processing requirement (representing the complexity of computation) and data processing rate (representing the amount of streams processed based on the speed of incoming streams). It can be either unmovable or movable. Unmovable service is a service that receives large amount of data that needs to be processed locally to avoid the time and cost of transferring data) such as vehicle detection service. Movable is a service that receives small working streams that can be transferred with low communication overhead of data transmission such as speed \& distance calculation service. Thus, with unmovable service, the data locality is applied to push service to the location of data while with movable service, placement optimization approach is applied to exploit the deployment flexibility.

The execution of the presented workflow is continuous. This means every service (e.g. roadside data analysis) receives continuous input streams from external source(s) (e.g. traffic light and camera) and/or internal source(s) (i.e. parent service(s) (e.g. speed and distance calculation), processes them continuously as they arrive and generate continuous output streams as the results of computations (e.g. pre-processed vehicle information), which routed towards one or more child services. The end service(s) such as traffic modelling generates the continuous output results for the execution of this workflow. In term of data mode that being used to route streams toward one or more child services, there are two modes: replica and partition. With replica mode, the output stream of parent service is replicated on child service(s). While with partition mode, the output stream of parent service is partitioned into portions based on the pre-defined partition percentages and then each portion is routed to the corresponding child service.

\begin{table}[H]
	\tiny
	\centering
	\caption{VM configurations of modelled clouds}
	\begin{tabular}{|p{4em}|c|p{2em}|p{5em}|c|c|p{4em}|}
		\hline
		Cloud Provider & VM Type& \parbox{0.5cm}{vCPUs/ cores} & ECUs & Total MIPS & Memory (GB) & Price ($\cent$/second) \\
		\hline
		\multirow{11}[1]{*}{\parbox{0.8cm}{Amazon EC2 (Windows instances)}} & m4.large & 2 & 6.5 (7) & 7000 & 8 & 0.0054 \\
		& m4.xlarge & 4 & 13 & 13000 & 16 & 0.0107 \\
		& m4.2xlarge & 8 & 26 & 26000 & 32 & 0.0214\\
		& m4.4xlarge & 16 & 53.5 (54) & 54000 & 64 & \\
		& m4.10xlarge & 40 & 124.5 (125) & 125000 & 160 & 0.1067 \\
		& m4.16xlarge & 64 & 188 & 188000 & 256 & 0.1707 \\
		& c4.large & 2 & 8 & 8000 & 3.75 & 0.0054 \\
		& c4.xlarge & 4 & 16 & 16000 & 7.5 & 0.0107 \\
		& c4.2xlarge & 8 & 31 & 31000 & 15 & 0.0213 \\
		& c4.4xlarge & 16 & 62 & 62000 & 30 & 0.0426 \\
		& c4.8xlarge & 36 & 132 & 132000 & 60 & 0.0859  \\
		\hline
		\multirow{13}[1]{*}{\parbox{0.8cm}{Google Compute Engine (n1-series)}} & n1-standard-1 & 1 & 2.75 & 2750 & 3.75 & 0.0014  \\
		& n1-standard-2 & 2 & 5.5 & 5500 & 7.5 & 0.0027 \\
		& n1-standard-4 & 4 & 11 & 11000 & 15 & 0.0053  \\
		& n1-standard-8 & 8 & 22 & 22000 & 30 & 0.0106  \\
		& n1-standard-16 & 16 & 44 & 44000 & 60 & 0.0212 \\
		& n1-standard-32 & 32 & 88 & 88000 & 120 & 0.0423 \\
		& n1-standard-64 & 64 & 176 & 176000 & 240 & 0.0845 \\
		& n1-highcpu-2 & 2 & 5.5 & 5500 & 1.8 & 0.002  \\
		& n1-highcpu-4 & 4 & 11 & 11000 & 3.6 & 0.004  \\
		& n1-highcpu-8 & 8 & 22 & 22000 & 7.2 & 0.0079  \\
		& n1-highcpu-16 & 16 & 44 & 44000 & 14.4 & 0.0158 \\
		& n1-highcpu-32 & 32 & 88 & 88000 & 28.8 & 0.0316 \\
		& n1-highcpu-64 & 64 & 176 & 176000 & 57.8 & 0.0631 \\
		\hline
		\multirow{16}[1]{*}{\parbox{0.8cm}{Microsoft Azure  (Windows D and F-Series)}} & D1 v2 & 1 & 2.5 & 2500 & 3.58 & 0.0035 \\
		& D2 v2 & 2 & 5 & 5000 & 7 & 0.0069 \\
		& D3 v2 & 4 & 10 & 10000 & 14 & 0.0137 \\
		& D4 v2 & 8 & 20 & 20000 & 28 & 0.0274 \\
		& D5 v2 & 16 & 40 & 40000 & 56 & 0.052 \\
		& D2 v3 & 2 & 5 & 5000 & 8 & 0.0054 \\
		& D4 v3 & 4 & 10 & 10000 & 16 & 0.0107 \\
		& D8 v3 & 8 & 20 & 20000 & 32 & 0.0214 \\
		& D16 v3 & 16 & 40 & 40000 & 64 & 0.0427 \\
		& D32 v3 & 32 & 80 & 80000 & 128 & 0.0854 \\
		& D64 v3 & 64 & 160 & 160000 & 256 & 0.1707 \\
		& F1 & 1 & 2.5 & 2500 & 2 & 0.0027 \\
		& F2 & 2 & 5 & 5000 & 4 & 0.0054 \\
		& F4 & 4 & 10 & 10000 & 8 & 0.0107 \\
		& F8 & 8 & 20 & 20000 & 16 & 0.0213 \\
		& F16 & 16 & 40 & 40000 & 32 & 0.0426 \\
		\hline

	\end{tabular}%
	\label{tab:VMConfigurations}%
\end{table}%

\subsection{Modelled Clouds and their VM Configurations}

To form a Multicloud environment for our experiments, we model three different cloud system providers, namely (Amazon EC2 \cite{Amazon2017Instances}, Google Cloud Engine \cite{Google2017Instances}, and Microsoft Azure \cite{Microsoft2017Instances}). Each cloud system has different VM configurations that chosen from pre-defined machine types offered by this cloud provider. Table \ref{tab:VMConfigurations} show the configurations of VM for modelled clouds. As IoTSim-Stream \cite{barika2019iotsim}, our proposed simulator that built on top of CloudSim to execute stream workflow applications in Multicloud environments, is used to run our experiments, the same computing power rating is applied. In CloudSim \cite{calheiros2011cloudsim}, MIPS rating is used to represent CPU unit, where the capacity of VM instance is represented by the total MIPS assigned to such instance based on the assigned value of MIPS rating multiplied by the number of assigned CPU cores (Processing Elements (PEs) in CloudSim term). Hence, the processing power of each VM instance offered by the modelled cloud is converted to the corresponding MIPS value.

\end{document}